\documentclass[aps,prl,reprint,superscriptaddress]{revtex4-2}
\usepackage{amsmath}
\usepackage{amsthm}

\usepackage{mathtools}
\usepackage{amssymb}
\usepackage{mathrsfs}
\usepackage[scr=boondox]   
{mathalpha}
\usepackage{graphicx}
\usepackage{dcolumn}
\usepackage{color}
\usepackage{ulem}
\usepackage{bm}
\usepackage{braket}
\usepackage[colorlinks,citecolor=blue]{hyperref}
\usepackage{multirow}
\usepackage{makecell}
\usepackage{booktabs}
\usepackage{graphicx}
\usepackage{booktabs}
\usepackage{multirow}
\usepackage{tabularx}
\usepackage{zref}
\usepackage{appendix}

\usepackage{lineno}

\usepackage{soul}

\newcommand\D{\mathrm{d}}
\newcommand\hrho{\hat{\rho}}
\newcommand\hrhos{\hat{\rho}_\text{ss}}
\newcommand\cs{C_\text{ss}}

\newcommand\sL{\mathscr{L}}

\newcommand\sA{\mathscr{a}}
\newcommand\sB{\mathscr{b}}

\newcommand\Le{L_\text{eff}}

\begin{document}

\title{Symmetry classification correspondence between quadratic Lindbladians and their steady states}

\author{Liang Mao}
\affiliation{Institute for Advanced Study, Tsinghua University, Beijing,100084, China}
\author{Fan Yang}
\email{Contact author: 101013867@seu.edu.cn}
\affiliation{School of Physics, Southeast University, No.2 SEU Road, Nanjing, China, 211189}

\date{\today}

\begin{abstract}
Symmetry classification is crucial in understanding universal properties of quantum matter. Recently, the scope of symmetry classification has been extended to open quantum systems governed by the Lindblad master equation. However, the classification of Lindbladians and steady states remains largely separate. Because the former requires the non-Hermitian classification framework, while the latter relies on the classification scheme for Hermitian matrices. In this paper we build connections between symmetry classes of quadratic Lindbladian and its steady state, despite their different classification frameworks. We classify the full matrix representation of generic quadratic Lindbladians with particle conservation, showing they fall into 27 non-Hermitian symmetry classes. Among these, 22 classes lead to an infinite-temperature steady state. The remaining five classes have one-to-one correspondence with five steady-state Hermitian symmetry classes. Numerical simulations of random Lindbladian dynamics confirm the convergence to the correct steady-state symmetry classes at long time.

\end{abstract}

\maketitle

Symmetry plays a significant role in understanding universal physical properties of quantum many-body systems. Many-body Hamiltonians are classified by the behavior under time-reversal, particle-hole, and chiral symmetries, leading to the celebrated Altland-Zirnbauer (AZ) tenfold way classification\cite{azclass1,azclass2,azclass3}. The AZ classification provides a powerful framework to understanding fermionic systems. This classification dictates the appearance of topological phases in translational-invariant gapped fermions based on symmetry class and dimensionality\cite{ti1,ti2,ti3}. Furthermore, in chaotic systems, the symmetry class determines universal random matrix correlations, defining level statistics\cite{disorder1,disorder2} and further the effective theory\cite{disorder3,disorder4}.

Recent studies of symmetry classification extend to open quantum systems, typically described by the Lindblad master equation\cite{open1,open2}
\begin{align}
	\frac{\D }{\D t}\hrho=\sL[\hrho] =-i[\hat{H},\hrho]+\sum_\mu \big(2\hat{L}_\mu\hrho\hat{L}_\mu^\dagger-\{\hat{L}_\mu^\dagger \hat{L}_\mu,\hrho  \} \big).
\end{align}
For such systems, both the long-time stationary state  and the transient dynamical phenomenon are of fundamental importance. The former is described by the steady state density matrix $\hrhos$. The symmetry and topology of $\hrhos$, particularly for fermionic Gaussian states amenable to AZ tenfold way classification, have been extensively investigated\cite{top_dens1,top_dens2,top_dens3,top_dens4,top_dens5,top_dens6,top_dens7,top_dens8}. Transient phenomena, on the other hand, is described by the dynamical generator $\sL$, dubbed the Lindbladian superoperator. When regarded as a linear map in the operator space, $\sL$ is in general non-Hermitian, preventing a direct AZ classification. Fortunately, the Bernard-LeClair (BL) classification for non-Hermitian matrices\cite{bl1}, recently well-developed and validated \cite{bl2,bl3,bl4,bl5,bl6}, offers a suitable framework. Applying the BL classification theory to $\sL$, several recent work unveiled universal topological\cite{open_top4,open_top5,open_top6,open_top1,open_top2,open_top7,open_top3} and quantum chaos\cite{open_chaos1,open_chaos2,open_chaos3,open_chaos4,open_chaos5,open_chaos6,open_chaos7,open_chaos8,open_chaos9,open_chaos10} properties of open quantum systems.

\begin{figure}[!b]
	\centering
	\includegraphics[width=0.45\textwidth]{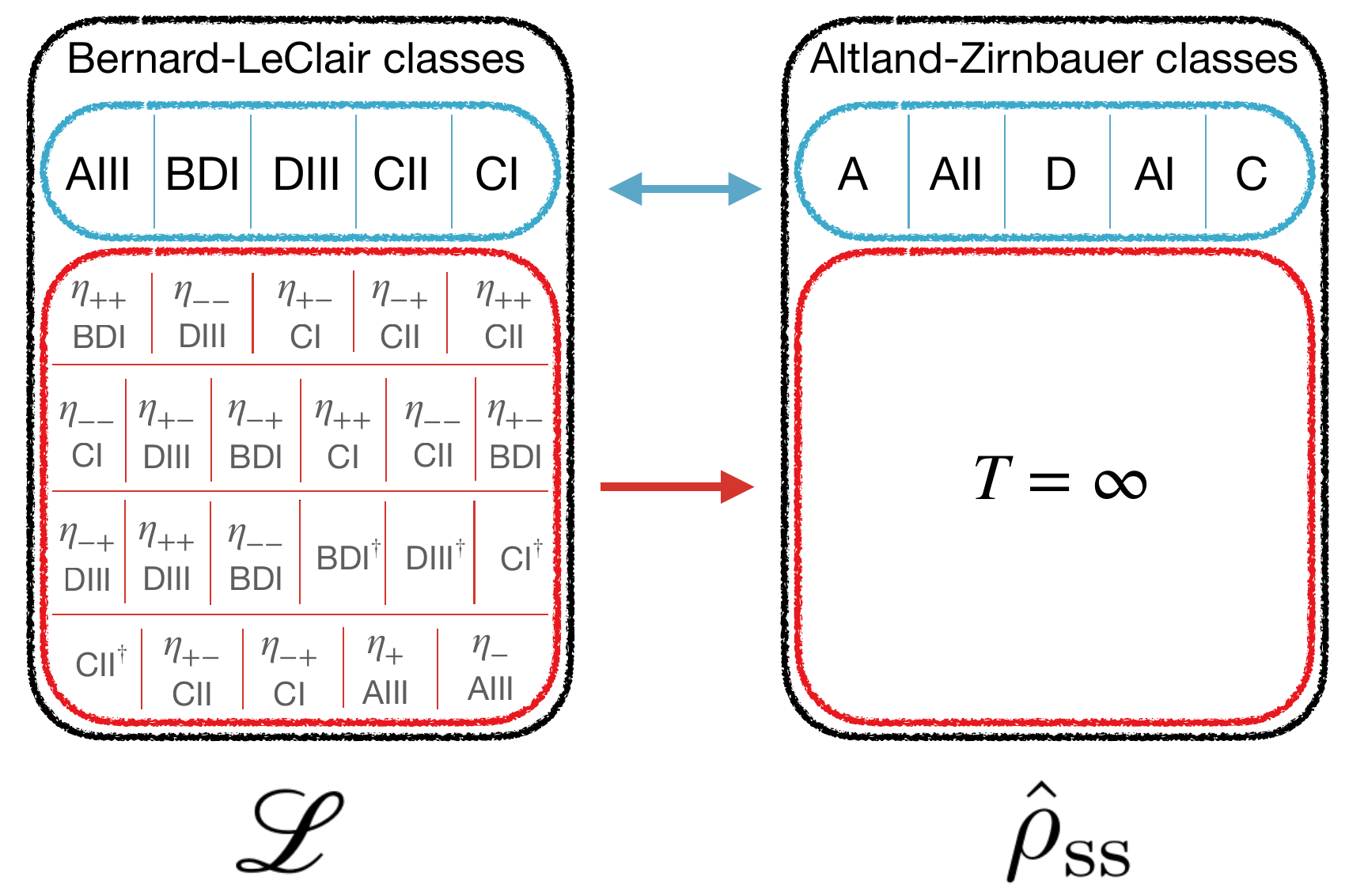}
	\caption{Symmetry classification correspondence between $\sL$ and $\hrhos$. Of 27 BL symmetry classes for $\sL$, 22 yield an infinite temperature steady state. The remaining 5 classes map one-to-one to 5 steady state AZ classes.  For definition of each class, see Tab.\ref{class}.}
	\label{illustration}
\end{figure}

However, a fundamental question still remains largely unexplored: \textit{How the symmetry classification of the Lindbladian affects the steady state, especially the symmetry classification of its steady state?} 
Given the physical intuition that transient dynamics gradually shape the steady state, we expect the symmetry classification of $\sL$ to dictate the physical properties of $\hrhos$. However, a gap arises when considering their symmetry classifications: $\sL$ and $\hrhos$ are described within different theoretical frameworks. Furthermore, the BL classification theory for $\sL$ comprises 54 possible symmetry classes\cite{bl2,bl3,bl4,bl5,bl6}, while the AZ classification for $\hrhos$ encompasses only ten. Due to this divergence,  the connections between symmetry classifications of $\sL$ and $\hrhos$ are yet to be clear.

This paper bridge the gap for quadratic fermionic Lindbladians with U(1) particle conservation. We formulate a precise correspondence between the BL symmetry classes of $\sL$ and the AZ symmetry classes of its unique steady state $\hrhos$. 
We extract the matrix representation $\Le$ of the Lindbladian by defining proper fermionic superoperators, a method dubbed third quantization\cite{3nd1,3nd2}.
Since $\sL$ must preserve trace and Hermiticity, $\Le$ is a highly structured matrix, allowing for 27 distinct BL symmetry classes. The structure of $\Le$ also constraints the form of symmetry transformations, and allows us to build connections to the steady state symmetries. Consequently, we further deduce that $\hrhos$ of 22 classes is the maximally mixed state. Each of the remaining 5 classes has a one-to-one correspondence with 5 of the 10 steady state AZ classes. The results are illustrated in Fig.\ref{illustration}

\section{ Symmetry classification overview.}
The focus of this paper is Lindbladians satisfying the following conditions.
 $\hat{H}=H_{ij}\hat{c}_i^\dagger \hat{c}_j$ is a quadratic operator of fermion operators $\hat{c}^\dagger$ and $\hat{c}$. Dissipation operators $\hat{L}_\mu$ are linear superpositions of either creation operators $\hat{L}_\mu=D^g_{\mu i}\hat{c}_i^\dagger$ or annihilation operators $\hat{L}_\mu=D^l_{\mu i}\hat{c}_i$. The total number of fermion modes is $N$. Thus $H$ is a $N\times N$ matrix. We also introduce $(M_l)_{ij}=D^{l*}_{\mu i}D^l_{\mu j}$ and $(M_g)_{ij}=D^{g*}_{\mu i}D^g_{\mu j}$ to characterize the structure of loss and gain. We note both $M_l$ and $M_g$ are positive semi-definite.

In the classification of topological insulators, the AZ classification theory is applied to the matrix representation $H$ of a quadratic Hamiltonian $\hat{H}$.
  For quadratic Lindbladians $\sL$, as pointed out in\cite{3nd1,3nd2}, we can extract a similar matrix representation. A quadratic Lindbladian can be written as a bilinear form using a set of fermionic superoperators, which we denote as $\sA$ and $\sB$, or more succinctly, super-fermions (SFs),
\begin{align}
	\sA_i[\hrho]\equiv \hat{c}_i\hrho,\quad
	\sB_i[\hrho]\equiv \mathcal{P}^F[\hrho]\hat{c}_i.
\end{align}
The corresponding SF creation superoperator is obtained through inner product $\mathrm{Tr}(\hrho_1^\dagger \sA[\hrho_2])^*=\mathrm{Tr}(\hrho_2^\dagger \sA^\dagger[\hrho_1])$, with $*$ denoting complex conjugation. 
Here $\mathcal{P}^F[\hrho]=(-1)^{\sum_i\hat{c}_i^\dagger\hat{c}_i}\hrho(-1)^{\sum_i\hat{c}_i^\dagger\hat{c}_i}$ is introduced to maintain the anti-commutation relation between two flavors of SFs,
\begin{align}
	\{\sA_i,\sA_j^\dagger\}=\delta_{ij},\quad 	\{\sB_i,\sB_j^\dagger\}=\delta_{ij}\notag\\
	\{\sA_i,\sA_j\}=\{\sB_i,\sB_j\}=\{\sA_i,\sB_j\}=0,
\end{align}
Physically, $\sA$ and $\sB$ are the fermions in the left and right subspaces after mapping the density matrix to a quantum state via Choi-Jamiolkowski isomorphism\cite{cj1,cj2}.

With SFs, we are able to write $\sL$ as a quadratic superoperator
\begin{align}\label{rep}
	\sL=&
    \begin{pmatrix}
    \sA^\dagger &\sB^\dagger
    \end{pmatrix}
    \begin{pmatrix}
		-iH-M_l+M_g^T & 2M_g^T\\ 2M_l & -iH+M_l-M_g^T
	\end{pmatrix}
    \begin{pmatrix}
    \sA \\\sB
    \end{pmatrix}
    \notag\\
    &+\text{Tr} (iH-M_l-M_g).
\end{align}
Here $\sA$ is the shorthand for the column vector $(\sA_1,\cdots,\sA_N)^T$, and the same  for $\sB$.
The constant term $\text{Tr}(\cdots)$ guarantees that the real parts of all the eigenvalues of $\sL$ are non-positive.
We denote the large $2N\times 2N$ matrix representation as $\Le$.
In this way, we translate the classification of superoperators $\sL$ to non-Hermitian matrix $L_\text{eff}$.

To classify $\Le$, we apply the BL classification framework for non-Hermitian matrices, where relevant symmetries are\cite{bl1,bl2,bl3,bl4,bl5}
\begin{align}\label{bl}
	\text{K sym:}&\quad \Le=\epsilon_K U_K\Le^*U_K^{-1},\quad U_KU_K^*=\eta_K\notag\\
	\text{C sym:}&\quad \Le=\epsilon_C U_C\Le^TU_C^{-1},\quad U_CU_C^*=\eta_C\notag\\
	\text{Q sym:}&\quad \Le=\epsilon_Q U_Q\Le^\dagger U_Q^{-1},\quad U_Q^2=1\notag\\
	\text{P sym:}&\quad \Le=-U_P\Le U_P^{-1},\quad U_P^2=1.
\end{align}
$U_{K,C,Q,P}$ are unitary matrices. $\epsilon_{K,C,Q}=\pm 1$, $\eta_{K,C}=\pm 1$ denote the sign of the symmetries. If more than one symmetry is present, symbols like $\epsilon_{PQ}=\pm 1$ are introduced to represent the commutation or anti-commutation relations between the symmetries. We denote  Q symmetry with $\epsilon_Q=\pm 1$ as the Q$_\pm$ symmetry, and other symmetries analogously. Counting the presence and sign of symmetries, there are 54 possible symmetry classes.

On the other hand, $\hrhos$ is a Hermitian operator. $\hrhos$ is characterized by the \textit{modular Hamiltonian} $\hat{G}$, such that $\hrhos=e^{-\hat{G}}/\mathcal{Z}$. For quadratic $\sL$, $\hat{G}$ is a quadratic operator $\hat{G}=G_{ij}\hat{c}_i^\dagger \hat{c}_j$. So its symmetry and topology can be understood from the conventional wisdom built from topological insulators. Symmetry classification of $G$ represent the classification of $\hrhos$\cite{top_dens1,top_dens2,top_dens3,top_dens4,top_dens5,top_dens6,top_dens7,top_dens8}.

For $G$, AZ symmetry classification is applied.
Hermiticity of $G$ makes some of the aforementioned symmetries identical. So the number of symmetries is reduced to three:
 time-reversal symmetry (TRS), particle-hole symmetry (PHS) and chiral symmetry (CS)\cite{azclass1,azclass2,azclass3},
\begin{align}
	\text{TRS:}&\quad G=V_TG^*V_T^{-1},\quad V_TV_T^*=\eta_T\notag\\
	\text{PHS:}&\quad G=-V_PG^TV_P^{-1},\quad V_PV_P^*=\eta_P\notag\\
	\text{CS:}&\quad G=-V_SGV_S^{-1},\quad V_S^2=1.
\end{align}
Here $V_{T,P,S}$ are unitary matrices. $\eta_T=\pm$ and $\eta_P=\pm$ label the sign of TRS and PHS. There are ten possible symmetry classes in total.

The goal of this work is to perform BL classification on $\Le$ and find its relation to AZ classification of $G$.

\begin{table*}[t!]
	
	\begin{tabularx}{0.92\textwidth}{>{\centering\arraybackslash}p{0.12\textwidth}>{\centering\arraybackslash}p{0.25\textwidth}>{\centering\arraybackslash}X}
		\toprule
		symmetry & transformation & constraints\\
		\toprule
        P&$\sigma_y\otimes V,\quad\sigma_z\otimes V$&
		$VHV^\dagger=-H,\quad VM_lV^\dagger=M_l\quad M_l=M_g^T$\\
		\midrule[0.3pt]
		C$_+$&$ \mathbb{I}_2\otimes V,\quad\sigma_x\otimes V$&
		$VHV^\dagger=H^T,\quad VM_lV^\dagger=M_l^T\quad M_l=M_g^T$\\
		\midrule[0.3pt]
		\multirow{2}{*}{C$_-$}&$\sigma_y\otimes V$&
		$VHV^{\dagger}=-H^T,\quad
		VM_lV^{\dagger} = M_l^T,\quad
		VM_g^TV^{\dagger}=M_g$\\
		&$\sigma_z\otimes V$&
		$VHV^{\dagger}=-H^T,\quad
		VM_lV^{\dagger} = M_g,\quad
		VM_g^TV^{\dagger}=M_l^T$\\
		\bottomrule
	\end{tabularx}

	\caption{Form of symmetry transformation and the constraints on $H$, $M_l$, and $M_g$ for P, C$_\pm$ symmetries.}
	\label{constraints}
\end{table*}

\section{ Summary of main results.}
We summarize the main results of this work as follows.

(1) \textbf{From 54 to 27}: A Lindbladian operator must retain the Hermiticity-preserving property. When $\sL$ operates on any $\hrho$, $\sL[\hrho]$ should always be a hermitian operator. This constraint is equivalent to a Q$_-$ symmetry $(\sigma_y\otimes \mathbb{I}_N)\Le (\sigma_y\otimes \mathbb{I}_N)^\dagger=-\Le^\dagger$. It eliminates 27 symmetry classes.

(2) \textbf{From 27 to 5}: Among the remaining 27 classes, there are 22 classes in which loss and gain terms share the same structure, leading to an infinite-temperature structureless steady state. Only Lindbladians in five symmetry classes can have a nontrivial steady state.

(3) \textbf{5 to 5 Correspondence}: There indeed exists a one-to-one correspondence between Lindbladians in the five remaining BL classes and steady-state density matrices in five AZ classes, as shown in Fig. \ref{illustration}.

A full list can be found in Tab.\ref{class} within Appendix. Below we present the derivation.

\section{ Derivation sketch.}

Building connections between $\Le$ and $G$ presents two main difficulties: (1) Although $\Le$ determines $G$, the concrete relation is complicated; (2) Symmetry transformations of $\Le$ and $G$ have different dimensions in their matrix representation. To overcome these, a crucial observation is that only $H$, $M_l$, and $M_g$ are free parameters  determining both $\Le$ and $G$. As a result, our strategy is to reduce both symmetry classifications of $\Le$ and $G$ to constraints on $H$, $M_l$ and $M_g$.

For $G$, we introduce the correlation matrix $(\cs)_{ij}=\text{Tr}(\hrhos\hat{c}^\dagger_i\hat{c}_j)$ to express $\hrhos$, since it is guaranteed to be Gaussian. $\cs$ is related to $G$ by $\cs=1/(e^{G^T}+1)$. Symmetries of $G$ translate directly to constraints of $\cs$, 
 \begin{align}\label{sym}
		\text{TRS}&:\quad V\cs V^\dagger -\cs^T=0,\notag\\
		\text{PHS}&:\quad V\cs V^\dagger +\cs^T-1=0,\notag\\
        \text{CS}&:\quad V\cs^T V^\dagger +\cs^T-1=0.
	\end{align}
Next, note that $\cs$ is determined by a linear equation\cite{eq1,eq2}
\begin{align}\label{ss}
	X\cs+\cs X^\dagger+2M_g=0.
\end{align}
where $X=iH^T-M_l^T-M_g$. When the equation has a unique solution (unique steady state), constraints of $\cs$ are further translated to constraints on $H$, $M_l$ and $M_g$.

For $\Le$, it is crucial to observe that the highly structured form of $\Le$ constrains its symmetry transformation $U$.  For generic parameters, we find $U$ takes the form $U=\sigma\otimes V$, where $V$ is a $N\times N$ unitary and $\sigma$ is a Pauli or identity matrix.  The concrete form of $\sigma$ varies among symmetries. The form of $U$ has a clear physical meaning. $V$ accounts for transformations inside each of the spaces of $\sA$ and $\sB$. $\sigma$ represents possible swaps between them. Since $\sA_i$ and $\sB_i$ are actually the same fermion mode $\hat{c}_i$ acting on different sides of density matrix, they should be transformed consistently by $V$. 
As a result, the symmetry transformation is reduced from $U$ to $V$, which can be further translated to constraints on $H$, $M_l$ and $M_g$.

\section{Symmetry transformation reduction.}
The constraints on $H$, $M_l$, and $M_g$ deduced from each symmetry are summarized in Tab.\ref{constraints}. Here we consider only P and C$_\pm$ symmetries, for a reason that will be clear later. The second column of the table shows the form of symmetry transformation $U$ of each symmetry. The third column shows the constraints on matrices $H$, $M_l$, and $M_g$.
A comprehensive derivation can be found in the Supplementary Materials. To illustrate the validity, we show two examples below

We first show a counterexample. Assume P symmetry is realized by $\sigma_x\otimes V$, violating the requirement of Table. \ref{constraints}. We then have
\begin{align}
	&\qquad (\sigma_x\otimes V)\Le(\sigma_x\otimes V)^\dagger\notag\\
	\small&=\begin{pmatrix}
		V(-iH+M_l-M_g^T )V^\dagger& 2VM_lV^\dagger\\
		2VM_g^TV^\dagger & V(-iH-M_l+M_g^T)V^\dagger
	\end{pmatrix}\notag\\
	&=-\Le.
\end{align}
This gives us $VM_lV^\dagger=-M_g^T$ by considering the $(1,2)$ off-diagonal block, contradicting the positive semi-definiteness of $M_l$ and $M_g$. It's clear that what matter are the form of $\Le$ and positive semi-definiteness of $M_{l,g}$.

\begin{table*}[t]
	\begin{tabularx}{0.9\textwidth}{>{\centering\arraybackslash}p{0.15\textwidth}>{\centering\arraybackslash}p{0.32\textwidth}>{\centering\arraybackslash}p{0.15\textwidth}>{\centering\arraybackslash}X}
		\toprule
		Class of $\Le$ & Symmetry of $\Le$& Class of $G$ & Symmetry of $G$\\
		\toprule
		AIII&Q$_-$&A&
		None\\
		BDI&Q$_-$ and C$_-$. $\eta_C=+$, $\epsilon_{QC}=+$&AII&
		TRS, $\eta_T=-$\\
		DIII&Q$_-$ and C$_-$. $\eta_C=+$, $\epsilon_{QC}=-$&D&
		PHS, $\eta_P=+$\\
		CII&Q$_-$ and C$_-$. $\eta_C=-$, $\epsilon_{QC}=+$&AI&
		TRS, $\eta_T=+$\\
		CI&Q$_-$ and C$_-$. $\eta_C=-$, $\epsilon_{QC}=-$&C&
		PHS, $\eta_P=-$\\
		\bottomrule
	\end{tabularx}

	\caption{Correspondence between BL class of $\Le$ and AZ class of $G$ for the five non-trivial classes.}
	\label{nontrivial}
\end{table*}

On the other hand, when P symmetry is realized by $\sigma_y\otimes V$, we have 
\begin{align}
	&\quad (\sigma_y\otimes V)\Le(\sigma_y\otimes V)^\dagger\notag\\
	\small&=\begin{pmatrix}
		V(-iH+M_l-M_g^T )V^\dagger& -2VM_lV^\dagger\\
		-2VM_g^TV^\dagger & V(-iH-M_l+M_g^T)V^\dagger
	\end{pmatrix}\notag\\
	&=
	-\Le.
\end{align}
Matching each of the four blocks, we get 
\begin{align}
    V(-iH+M_l-M_g^T )V^\dagger&=iH+M_l-M_g^T,\notag\\
    VM_lV^\dagger=M_g^T,&\quad VM_g^TV^\dagger=M_l.
\end{align}
It then gives
\begin{align}
	VHV^\dagger=-H,\quad VM_lV^\dagger=M_g^T,\quad M_l=M_g^T.
\end{align}
This verifies $\sigma_y\otimes V$ as a legitimate P symmetry transformation. Moreover, the constraints on $H$, $M_l$, and $M_g$ are consistent with Tab.\ref{constraints}.

\section{ Classification of $\Le$ and steady state properties}
Now we analyze the BL symmetry classification of $\Le$, and discuss the steady state properties. We assume that $\sL$ has a unique steady state.
The full results are summarized comprehensively in Tab.\ref{class}.

First of all, as previously mentioned, $\Le$ has an inherent Q$_-$ symmetry expressed as $(\sigma_y\otimes \mathbb{I}_N)\Le (\sigma_y\otimes \mathbb{I}_N)^\dagger=-\Le^\dagger$. There are only 27 symmetry classes including this symmetry, which are therefore suitable for $\Le$. With the Q$_-$ symmetry present, K and C symmetries become indistinguishable. So only P and C$_\pm$ symmetries need to be considered for subsequent classification. That is the reason why we only consider them before.

Next, from Tab. \ref{constraints}, we could find that
when P or C$_+$ symmetries are present, they both require $M_l=M_g^T$. This constraint means that gain and loss have the same structure. Such kind of dissipation heats the steady state to an infinite temperature state.
From Eq.\ref{ss}, we can find when $M_l=M_g^T$, the solution is $\cs=\mathbb{I}_N/2$, corresponding to an infinite temperature state.
There are 22 symmetry classes containing at least one of P and C$_+$ symmetries (Tab.\ref{class}). So they all have infinite temperature state as the steady state. We note that when $M_l=M_g^T$, $\Le$ can be brought into block-diagonal form $\text{diag}\{-X^T,X^*\}$. So the symmetry class can also be defined by symmetries of $X^T$. See EM for detailed definition of each class.

Finally, the 5 BL classes remaining are: AIII, BDI, DIII, CII and CI, summarized in Tab.\ref{nontrivial}. Class AIII Lindbladian contains only the inherent Q$_-$ symmetry without any other constraints. So its steady state also have no symmetry constraints, thus belonging to the AZ class A. 
Class BDI and CII both have C$_-$ symmetry with $\epsilon_{QC}=+$. This C$_-$ symmetry is realized by $\sigma_y\otimes V$ (Tab.\ref{constraints}). Similarly, the C$_-$ symmetry of class DIII and CI is realized by $\sigma_z\otimes V$. 
In the following, we show that the constraints of these two types of symmetries lead steady states with TRS and PHS, respectively.

Consider the case that $\Le$ has C$_-$ symmetry realized by $\sigma_y\otimes V$. We take the transpose of Eq. \ref{ss} and substitute in the expression of $H$, $M_l$ and $M_g^T$ (Tab.\ref{constraints}), which gives
\begin{align}
	X(V^\dagger \cs^T V )+(V^\dagger\cs^T V )X^\dagger +2M_g=0.
\end{align}
For general coefficients, the above equation is consistent with Eq.\ref{ss} only when $V\cs V^\dagger =\cs^T$. Then from Eq.\ref{sym},
we find that $G$ has TRS.

Crucially, this is a necessary-sufficient condition. Assume the unique steady state $G$ has TRS.
We can also take the transpose of Eq. \ref{ss} and plugging the expression of $\cs^T$ (Eq.\ref{sym}) into it. It then becomes
\begin{align}
	(V^\dagger X^*V)\cs +\cs (V^\dagger X^T V) +2 V^\dagger M_g^T V=0.
\end{align} 
Matching the coefficients with Eq. \ref{ss}, we have
\begin{align}
	V^\dagger X^*V=X,V^\dagger X^T V=X^\dagger, V^\dagger M_g^T V=M_g.
\end{align}
These equations recover the constraints in Tab. \ref{constraints} for C$_-$ symmetry with transformation $\sigma_y\otimes V$. 
Thus, the C$_-$ symmetry of $\Le$ realized by $\sigma_y\otimes V$ and TRS of $G$ are equivalent.

For $\Le$ with C$_-$ symmetry realized by $\sigma_z\otimes V$, its equivalence to steady state with PHS follows analogously. Similar discussions have been carried out in\cite{open_top2,open_top3}.

Based on the above results, we can establish the correspondence between the remaining four BL classes and four of the AZ classes as the follows. When $\Le$ is in the BL class BDI, it has C$-$ symmetry realized by $\sigma_y\otimes V$, and $\eta_C=+$. So $VV^*=-1$. Thus, the steady state must have TRS with $\eta_T=-$, i.e., must be in the AZ class AII. $\Le$ in BL class CII also has C$-$ symmetry realized by $\sigma_y\otimes V$, but $\eta_C=-$. So the steady state must have TRS with $\eta_T=+$, thus must be in the AZ class AI. Similarly, the BL classes DIII and CI correspond to the AZ classes D and C, respectively.

In summary, $\Le$ in BL classes AIII, BDI, DIII, CII and CI correspond to $G$ belonging to AZ classes A, AII, D, AI and C, respectively (Tab.\ref{nontrivial}). The five steady state classes here are the ones that do not possess chiral symmetry.  When steady state has chiral symmetry from the combination of TRS and PHS, $\Le$ has two different C$_-$ symmetries $U_1=\sigma_y\otimes V_T$ and $U_2=\sigma_z\otimes V_P$. They are combined to a commuting unitary symmetry $(U_1U_2^\dagger)\Le(U_1U_2^\dagger)^\dagger=-U_1\Le^T U_1^\dagger=\Le$, whose transformation is $U=U_1U_2^\dagger=\sigma_x\otimes V_S$. Since $V_S^2=1$, we have $U^2=1$\footnote{We can always achieve this goal by properly add a phase factor $e^{i\theta}$ to $U$.}. So $\Le$ has the block-diagonal form $\Le^+\oplus\Le^-$ defined by  $U=\pm 1$ . For a comprehensive  classification, one need to study each block $\Le^{\pm}=(U\pm \mathbb{I}_{2N})\Le/2$, where $V_S$ is involved in the classification procedure. Moreover, the inherent Q$_-$ symmetry is broken down to a relation $(\sigma_y\otimes \mathbb{I}_{N})L_\text{eff}^+(\sigma_y\otimes \mathbb{I}_N)^\dagger=(L_\text{eff}^-)^\dagger$. So the structure of $L_\text{eff}$ is lost. We leave these more complicated cases involving chiral symmetry for future studies.

\begin{figure}[!t]
	\centering
	\includegraphics[width=0.48\textwidth]{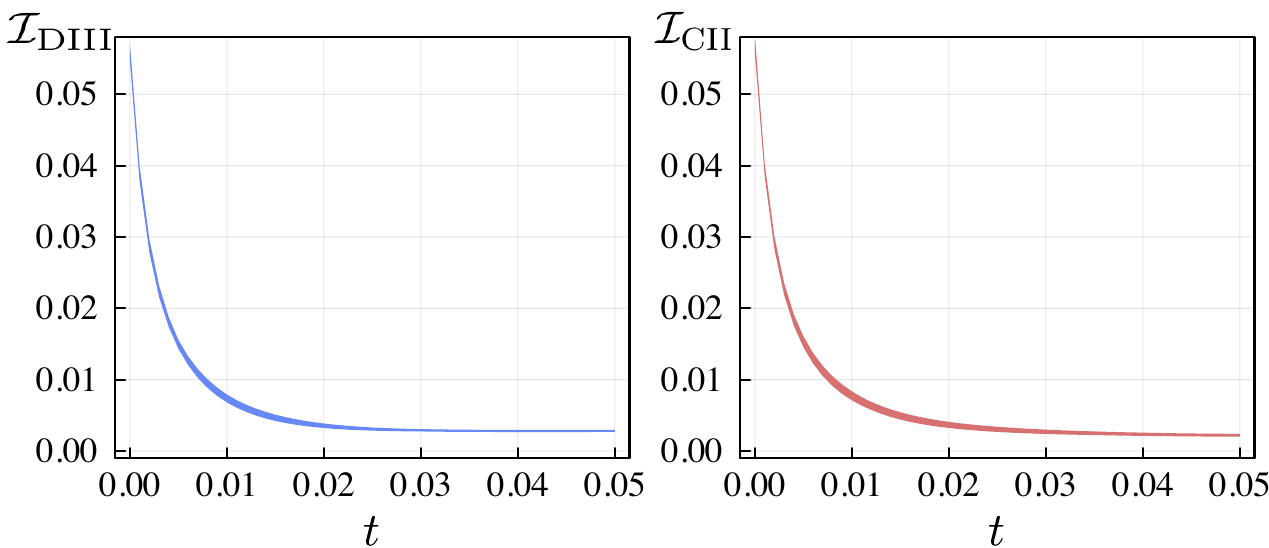}
	\caption{Dynamics of symmetry indicators of {\bf Left:} $\mathrm{DIII}$ class and {\bf Right:} $\mathrm{CII}$ class over 100 samplings. The boundaries of shaded region represents the maximum and minimum values that symmetry indicators can achieve at each time by the samplings. A good convergence to desired AZ symmetry classes at late time is shown, with the universal convergence dynamics.}
	\label{dynamics}
\end{figure}

\section{Convergence dynamics of random Lindbladians}
To illustrate the validity of our results, we numerically simulate the dynamics of random Lindbladians belonging to BL classes DIII and CII with a random initial state\footnote{The codes for numerical calculation are available at https://github.com/maol17/2503.10763}. We show that the state converges to AZ classes D and AI at late time.

For simplicity, we choose $V=\text{diag}\{1,-1,1,-1,\cdots\}$ with $N$ being even. We sample each matrix entry of $H$, $M_l$, and $M_g$ from Gaussian distribution $\mathcal{N}(0,1)$, while retaining the constraints shown in Tab. \ref{constraints}. We sample initial state from the ground state of a random parent Hamiltonian, with matrix entries again sampled from $\mathcal{N}(0,1)$.

To quantify how much a given state approaches the desired symmetry classes, we use the following symmetry indicators of TRS and PHS
\begin{align}
	\mathcal{I}_\text{DIII} &= \Vert VCV^\dagger -C^T \Vert_1/N^2,\notag\\
	\mathcal{I}_\text{CII} &= \Vert VCV^\dagger +C^T-1 \Vert_1/N^2.
\end{align}
Here $C$ is the correlation matrix of the state, and $\Vert\cdots\Vert_1$ denotes the trace norm.
We plot the trajectories of symmetry indicators over 100 individual random samplings of both Lindblad parameters and initial states. Each trajectory exhibits good convergence to the desired symmetry classes at late times, verifying the validity of our results.

\section{ Summary and discussion.}
In this paper, we establish a symmetry classification correspondence between quadratic fermionic Lindbladian and its steady state. We perform a BL symmetry classification on the Lindbladian and find 27 possible symmetry classes.
 In these classes, 22 of them lead to an infinite temperature steady state. The remaining 5 classes have a one-to-one correspondence with 5 steady state AZ classes.

Our findings imply that universal physical properties of Lindbladian and steady state are connected. For example, when Lindbladian spectrum has a certain universal level statistics, the steady state level statistics is also fixed by the symmetry classification correspondence. The connection has been partially found in\cite{random3,random2}. Understanding the physical mechanism of such connections for other symmetry classes could be an interesting direction. Finding connections of topological properties could also be intriguing. These studies are helpful to understand the mechanism of symmetry classification correspondence.

Our results potentially benefit state-preparation experiments by offering guidance for the symmetry constraints of dissipation channels.
As an example, to simulate the quantum spin hall effect with, e.g., the Kane-Mele model\cite{kane2005z,open_top4} (class AII, two dimensions), the Lindbladian need to be in BDI class. 
For the purpose of dissipative engineering, one should engineer $H$ to belong to class C, and $M_l$, $M_g$ to belong to class AII. For nearly dissipation-free systems to simulate topological dynamics, instead, TRS-breaking dissipation is the dominant channel thus should be minimized.

While applied to the $U(1)$-symmetric cases, our framework can be directly generalized to those without $U(1)$ symmetry. A generic quadratic Lindbladian takes a Bogoliubov-de Gennes (BdG) form of matrix representation $\sL=\Psi^\dagger\Le'\Psi+\text{const}$, where $\Psi=(\sA,\sB,\sA^\dagger,\sB^\dagger)^T$. The steady state modular Hamiltonian also takes the BdG form $\hat{G}=\Phi^\dagger G'\Phi$, with $\Phi=(\hat{c},\hat{c}^\dagger)^T$. A similar method can be used to study the relation of $\Le'$ and $G'$. 

Furthermore, by incorporating the self-energy of modular Hamiltonian and the self-energy of Lindbladians \cite{Wang24}, it is possible to generally study the symmetry classification correspondence for interacting fermionic systems. We leave this to future work.

\section{Acknowledgments}
We thank Hui Zhai, Zhong Wang, He-Ran Wang, Zijian Wang, Fei Song, Tian-Shu Deng, and Sebastian Diehl for helpful discussions. This work is supported by the National Key R$\&$D Program of China (Grant No. 2023YFA1406702) and the Start-up Research Fund of Southeast University (RF1028624190).

\begin{table*}
	\begin{tabularx}{0.99\textwidth}{>{\centering\arraybackslash}p{0.1\textwidth}>{\centering\arraybackslash}p{0.46\textwidth}>{\centering\arraybackslash}p{0.15\textwidth}>{\centering\arraybackslash}X}
		\toprule[1pt]
		Class of $\Le$ & Symmetry of $\Le$& $-X^T\oplus X^*$&Steady state\\
		\toprule[1pt]
		AIII&Q$_-$&\multirow{5}{*}{Not applicable}&
		AZ class A, T=0, P=0, S=0\\
		BDI&Q$_-$ and C$_-$. $\eta_C=+$, $\epsilon_{QC}=+$&&
		AZ class AII, T=$-$, P=0, S=0\\
		DIII&Q$_-$ and C$_-$. $\eta_C=+$, $\epsilon_{QC}=-$&&
		AZ class D, T=0, P=+, S=0\\
		CII&Q$_-$ and C$_-$. $\eta_C=-$, $\epsilon_{QC}=+$&&
		AZ class AI, T=$+$, P=0, S=0\\
		CI&Q$_-$ and C$_-$. $\eta_C=-$, $\epsilon_{QC}=-$&&
		AZ class C, T=0, P=$-$, S=0\\
		\midrule[0.5pt]
		BDI$^\dagger$&Q$_-$ and C$_+$. $\eta_C=+$, $\epsilon_{QC}=+$&\multirow{2}{*}{\makecell{AI$^\dagger$$\oplus$AI$^\dagger$\\ C$_+$, $\eta_C=+$}}&
		\multirow{22}{*}{Infinite $T$}\\
		CI$^\dagger$&Q$_-$ and C$_+$. $\eta_C=+$, $\epsilon_{QC}=-$&&\\
        \cmidrule(lr){1-3}
		CII$^\dagger$&Q$_-$ and C$_+$. $\eta_C=-$, $\epsilon_{QC}=+$&\multirow{2}{*}{\makecell{AII$^\dagger$$\oplus$AII$^\dagger$\\ C$_+$, $\eta_C=-$}}&\\
		DIII$^\dagger$&Q$_-$ and C$_+$. $\eta_C=-$, $\epsilon_{QC}=-$&&\\
        \cmidrule(lr){1-3}
		$\eta_+$AIII&Q$_-$ and P. $\epsilon_{PQ}=+$&\multirow{2}{*}{$\eta$A$\oplus$$\eta$A, Q$_+$}&\\
		$\eta_-$AIII&Q$_-$ and P. $\epsilon_{PQ}=-$&&\\
        \cmidrule(lr){1-3}
		$\eta_{++}$BDI&Q$_-$, P and C$_+$. $\eta_C=+$ $\epsilon_{QC}=+$, $\epsilon_{PQ}=+$ $\epsilon_{PC}=+$&\multirow{4}{*}{\makecell{$\eta_+$AI$\oplus$$\eta_+$AI\\ Q$_+$, C$_+$\\ $\eta_C=+$, $\epsilon_{QC}=+$}}&\\
        $\eta_{-+}$CII&Q$_-$, P and C$_+$. $\eta_C=+$ $\epsilon_{QC}=-$, $\epsilon_{PQ}=-$ $\epsilon_{PC}=-$&&\\
        $\eta_{++}$CI&Q$_-$, P and C$_+$. $\eta_C=+$ $\epsilon_{QC}=+$, $\epsilon_{PQ}=-$ $\epsilon_{PC}=+$&&\\
        $\eta_{-+}$DIII&Q$_-$, P and C$_+$. $\eta_C=+$ $\epsilon_{QC}=-$, $\epsilon_{PQ}=+$ $\epsilon_{PC}=-$&&\\
        \cmidrule(lr){1-3}
		$\eta_{--}$DIII&Q$_-$, P and C$_+$. $\eta_C=+$ $\epsilon_{QC}=+$, $\epsilon_{PQ}=+$ $\epsilon_{PC}=-$&\multirow{4}{*}{\makecell{$\eta_-$AII$\oplus$$\eta_-$AII\\ Q$_+$, C$_+$\\ $\eta_C=+$, $\epsilon_{QC}=-$}}&\\
		$\eta_{+-}$CI&Q$_-$, P and C$_+$. $\eta_C=+$ $\epsilon_{QC}=-$, $\epsilon_{PQ}=-$ $\epsilon_{PC}=+$&&\\
        $\eta_{--}$CII&Q$_-$, P and C$_+$. $\eta_C=+$ $\epsilon_{QC}=+$, $\epsilon_{PQ}=-$ $\epsilon_{PC}=-$&&\\
        $\eta_{+-}$BDI&Q$_-$, P and C$_+$. $\eta_C=+$ $\epsilon_{QC}=-$, $\epsilon_{PQ}=+$ $\epsilon_{PC}=+$&&\\
        \cmidrule(lr){1-3}
		$\eta_{++}$CII&Q$_-$, P and C$_+$. $\eta_C=-$ $\epsilon_{QC}=+$, $\epsilon_{PQ}=+$ $\epsilon_{PC}=+$&\multirow{4}{*}{\makecell{$\eta_+$AII$\oplus$$\eta_+$AII\\ Q$_+$, C$_+$\\ $\eta_C=-$, $\epsilon_{QC}=+$}}&\\
        $\eta_{-+}$BDI&Q$_-$, P and C$_+$. $\eta_C=-$ $\epsilon_{QC}=-$, $\epsilon_{PQ}=-$ $\epsilon_{PC}=-$&&\\
        $\eta_{++}$DIII&Q$_-$, P and C$_+$. $\eta_C=-$ $\epsilon_{QC}=+$, $\epsilon_{PQ}=-$ $\epsilon_{PC}=+$&&\\
        $\eta_{-+}$CI&Q$_-$, P and C$_+$. $\eta_C=-$ $\epsilon_{QC}=-$, $\epsilon_{PQ}=+$ $\epsilon_{PC}=-$&&\\
        \cmidrule(lr){1-3}
		$\eta_{--}$CI&Q$_-$, P and C$_+$. $\eta_C=-$ $\epsilon_{QC}=+$, $\epsilon_{PQ}=+$ $\epsilon_{PC}=-$&\multirow{4}{*}{\makecell{$\eta_-$AI$\oplus$$\eta_-$AI\\ Q$_+$, C$_+$\\ $\eta_C=-$, $\epsilon_{QC}=-$}}&\\
		$\eta_{+-}$DIII&Q$_-$, P and C$_+$. $\eta_C=-$ $\epsilon_{QC}=-$, $\epsilon_{PQ}=-$ $\epsilon_{PC}=+$&&\\
		$\eta_{--}$BDI&Q$_-$, P and C$_+$. $\eta_C=-$ $\epsilon_{QC}=+$, $\epsilon_{PQ}=-$ $\epsilon_{PC}=-$&&\\
		$\eta_{+-}$CII&Q$_-$, P and C$_+$. $\eta_C=-$ $\epsilon_{QC}=-$, $\epsilon_{PQ}=+$ $\epsilon_{PC}=+$&&\\
		\bottomrule[1pt]
	\end{tabularx}

	\caption{All possible $\Le$ symmetry classes and their relation to steady state. {\bf Column one}: symmetry class of $\Le$, defined by column two. {\bf Column two}: defining symmetries of each class.  {\bf Column three}: symmetry class of $-X^T\oplus X^*$, when $M_l=M_g^T$.  {\bf Column four}: Steady state properties.}
	\label{class}
\end{table*}
\section{Appendix}

\subsection{Definition and construction of symmetry class}
For a detailed summary of results, see Tab.\ref{class}. 
The class of $\Le$ is primarily defined by its symmetries, which is shown in the first and second columns of the table. For the name of each class, see\cite{bl3,bl5}.

In the last 22 classes, however, $M_l=M_g^T$ holds. At this time, $\Le$ can be transformed into block-diagonal form $U\Le U^\dagger=\text{diag}\{-X^T,X^*\}$ (see Eq.\ref{dynamics} for definition of $X$), for $U=(\sigma_y+\sigma_z)\otimes \mathbb{I}_N/\sqrt{2}$. So the classification of $\Le$ can also be expressed by symmetries of its diagonal block $X^T$. We show the class of $-X^T\oplus X^*$ defined in this way in the third column of Tab.\ref{class}. If we count only different classes of the diagonal block, the number of distinct classes is reduced from 22 to 7.

To see how each class is constructed, we take class $\eta_{++}$BDI as a representative example. In this class, $\epsilon_Q=+$ and $\epsilon_{QC}=+$. While the inherent Q symmetry is $\sigma_y\otimes \mathbb{I}_N$, C$_+$ symmetry should be realized by $\mathbb{I}_2\otimes V_C$ (Tab.\ref{constraints}). Similarly, P symmetry should be realized by $\sigma_y\otimes V_P$. Because of $\epsilon_{PC}=+$, we further demand $V_PV_C=V_CV_P$. Since $\eta_C=+$, we have $V_CV_C^*=1$.

In this case, we have $V_PX^TV_P^\dagger=(X^T)^\dagger$ and $V_CX^TV_C^\dagger=(X^T)^T$. So $X^T$ has Q$_+$ and C$_+$ symmetries with $\epsilon_{QC}=+$, $\eta_C=+$. As a result, $-X^T\oplus X^*$ belongs to class $\eta_+$AI$\oplus$$\eta_+$AI.

The above construction can be carried out for all the 27 classes.

\subsection{Derivation of symmetry transformation reduction}
To begin with, we rewrite $\Le$ as
\begin{align}\label{form}
	\Le&=
\mathbb{I}_2\otimes(-iH)+\sigma_z\otimes (M_g^T-M_l)+\notag\\
&\sigma_x\otimes (M_l+M_g^T)+i\sigma_y\otimes(M_g^T-M_l).
\end{align}
It is natural to introduce new free parameters $M_+=M_l+M_g^T$, $M_-=M_g^T-M_l$. The former represents total dissipation and the latter characterize fluctuations\cite{open_top2}. We assume the coefficients are general. Matrices $H$, $M_+$ and $M_-$ are independent and could possess some symmetries themselves.

By the form of $\Le$, we should consider symmetry transformation that takes the form $U=u\otimes V$, where $u$ is a $2\times 2$ unitary and $V$ is a $N\times N$ unitary matrix. Physically, $V$ acts inside the spaces of each flavor $\sA$ and $\sB$. 
Recall that $\sA$ and $\sB$ are just the same fermion mode acting on different sides of the density matrix.
They are subject to two physical constraints: (1) the flavors $\sA$ and $\sB$ should either swap or not. (2) The transformations inside $\sA$ and $\sB$ should be the same up to a global phase. As a result, $u$ should take the form of either $\cos\theta\mathbb{I}_2+i\sin\theta\sigma_z$ or $\cos\phi\sigma_x+\sin\phi\sigma_y$ .

For the P symmetry, the matrix equation is
	\begin{gather}
		\mathbb{I}_2\otimes(-iH)+
		\sigma_x\otimes M_++(\sigma_z+i\sigma_y)\otimes M_-=\notag\\
		-\mathbb{I}_2\otimes(-iVHV^\dagger)- 
		u\sigma_xu^\dagger\otimes
		VM_+V^\dagger\notag\\
        -u(\sigma_z+i\sigma_y)u^\dagger\otimes VM_-V^\dagger.
	\end{gather}
The three terms should all match each other. The first term gives $VHV^\dagger=-H$.  
As for the second term, note that $M_+=M_l+M_g^T$ is a positive semi-definite matrix. So we can only have $u\sigma_x u^\dagger=-\sigma_x$, $VM_+V^\dagger=M_+$. This gives $u=\sigma_y$ or $\sigma_z$. For the third term, we can have $u(\sigma_z+i\sigma_y)u^\dagger=\pm(\sigma_z+i\sigma_y)$. However, none of the them can be realized by the former choice of $u$. As a result, coefficient $M_-$ must vanish (Recall that $M_+=M_l+M_g^T$, so it cannot vanish). Together, these give $M_l=M_g^T$, $VM_lV^\dagger=M_l$, and $u=\sigma_z$ or $u=\sigma_y$.

For the C$_+$ symmetry, the matrix equation is
	\begin{gather}
	\mathbb{I}_2\otimes(-iH^T)+
		\sigma_x\otimes M_+^T+(\sigma_z-i\sigma_y)\otimes M_-^T=\notag\\
		\mathbb{I}_2\otimes(-iVHV^\dagger)+ 
		u\sigma_xu^\dagger\otimes
		VM_+V^\dagger\notag\\
        +u(\sigma_z+i\sigma_y)u^\dagger\otimes VM_-V^\dagger.
\end{gather}
The first term gives $VHV^\dagger=H^T$. The second terms gives $u\sigma_x u^\dagger=\sigma_x$, $VM_+V^\dagger=M_+^T$. So $u=\mathbb{I}_2$ or $\sigma_x$. But these $u$ cannot realize $u(\sigma_z+i\sigma_y)u^\dagger=\pm(\sigma_z-i\sigma_y)$. As a result, we also have $M_-=0$. Together these give $M_l=M_g^T$, $VM_lV^\dagger=M_l^T$, and $u=\mathbb{I}_2$ or $u=\sigma_x$.

For the C$_-$ symmetry, the matrix equation is
\begin{gather}
	\mathbb{I}_2\otimes(-iH^T)+
		\sigma_x\otimes M_+^T+(\sigma_z-i\sigma_y)\otimes M_-^T=\notag\\
		-\mathbb{I}_2\otimes(-iVHV^\dagger)- 
		u\sigma_xu^\dagger\otimes
		VM_+V^\dagger\notag\\
        -u(\sigma_z+i\sigma_y)u^\dagger\otimes VM_-V^\dagger.
\end{gather}
The first term gives $VHV^\dagger=-H^T$. The second term gives $VM_+V^\dagger=M_+^T$, $u\sigma_xu^\dagger=-\sigma_x$. So $u=\sigma_y$ or $\sigma_z$. The third term requires $u(\sigma_z+i\sigma_y)u^\dagger=\pm(\sigma_z-i\sigma_y)$. It can be realized by either $u=\sigma_y$ or $u=\sigma_z$. For the former, $u(\sigma_z+i\sigma_y)u^\dagger=-\sigma_z+i\sigma_y$, $VM_-V^\dagger=M_-^T$. Together with $VM_+V^\dagger=M_+^T$, we have 
\begin{align}
    VM_lV^\dagger=M_l^T,\quad VM_g^TV^\dagger=M_g.
\end{align}
For the latter, $u(\sigma_z+i\sigma_y)u^\dagger=\sigma_z-i\sigma_y$, $VM_-V^\dagger=-M_-^T$. It gives
\begin{align}
    VM_lV^\dagger=M_g,\quad VM_g^TV^\dagger=M_l^T.
\end{align}

\bibliography{draftNotes}

\begin{thebibliography}{55}%
\makeatletter
\providecommand \@ifxundefined [1]{%
 \@ifx{#1\undefined}
}%
\providecommand \@ifnum [1]{%
 \ifnum #1\expandafter \@firstoftwo
 \else \expandafter \@secondoftwo
 \fi
}%
\providecommand \@ifx [1]{%
 \ifx #1\expandafter \@firstoftwo
 \else \expandafter \@secondoftwo
 \fi
}%
\providecommand \natexlab [1]{#1}%
\providecommand \enquote  [1]{``#1''}%
\providecommand \bibnamefont  [1]{#1}%
\providecommand \bibfnamefont [1]{#1}%
\providecommand \citenamefont [1]{#1}%
\providecommand \href@noop [0]{\@secondoftwo}%
\providecommand \href [0]{\begingroup \@sanitize@url \@href}%
\providecommand \@href[1]{\@@startlink{#1}\@@href}%
\providecommand \@@href[1]{\endgroup#1\@@endlink}%
\providecommand \@sanitize@url [0]{\catcode `\\12\catcode `\$12\catcode
  `\&12\catcode `\#12\catcode `\^12\catcode `\_12\catcode `\%12\relax}%
\providecommand \@@startlink[1]{}%
\providecommand \@@endlink[0]{}%
\providecommand \url  [0]{\begingroup\@sanitize@url \@url }%
\providecommand \@url [1]{\endgroup\@href {#1}{\urlprefix }}%
\providecommand \urlprefix  [0]{URL }%
\providecommand \Eprint [0]{\href }%
\providecommand \doibase [0]{https://doi.org/}%
\providecommand \selectlanguage [0]{\@gobble}%
\providecommand \bibinfo  [0]{\@secondoftwo}%
\providecommand \bibfield  [0]{\@secondoftwo}%
\providecommand \translation [1]{[#1]}%
\providecommand \BibitemOpen [0]{}%
\providecommand \bibitemStop [0]{}%
\providecommand \bibitemNoStop [0]{.\EOS\space}%
\providecommand \EOS [0]{\spacefactor3000\relax}%
\providecommand \BibitemShut  [1]{\csname bibitem#1\endcsname}%
\let\auto@bib@innerbib\@empty
\bibitem [{\citenamefont {Altland}\ and\ \citenamefont
  {Zirnbauer}(1997)}]{azclass1}%
  \BibitemOpen
  \bibfield  {author} {\bibinfo {author} {\bibfnamefont {A.}~\bibnamefont
  {Altland}}\ and\ \bibinfo {author} {\bibfnamefont {M.~R.}\ \bibnamefont
  {Zirnbauer}},\ }\bibfield  {title} {\bibinfo {title} {Nonstandard symmetry
  classes in mesoscopic normal-superconducting hybrid structures},\ }\href@noop
  {} {\bibfield  {journal} {\bibinfo  {journal} {Physical Review B}\ }\textbf
  {\bibinfo {volume} {55}},\ \bibinfo {pages} {1142} (\bibinfo {year}
  {1997})}\BibitemShut {NoStop}%
\bibitem [{\citenamefont {Schnyder}\ \emph {et~al.}(2008)\citenamefont
  {Schnyder}, \citenamefont {Ryu}, \citenamefont {Furusaki},\ and\
  \citenamefont {Ludwig}}]{azclass2}%
  \BibitemOpen
  \bibfield  {author} {\bibinfo {author} {\bibfnamefont {A.~P.}\ \bibnamefont
  {Schnyder}}, \bibinfo {author} {\bibfnamefont {S.}~\bibnamefont {Ryu}},
  \bibinfo {author} {\bibfnamefont {A.}~\bibnamefont {Furusaki}},\ and\
  \bibinfo {author} {\bibfnamefont {A.~W.}\ \bibnamefont {Ludwig}},\ }\bibfield
   {title} {\bibinfo {title} {Classification of topological insulators and
  superconductors in three spatial dimensions},\ }\href@noop {} {\bibfield
  {journal} {\bibinfo  {journal} {Physical Review B—Condensed Matter and
  Materials Physics}\ }\textbf {\bibinfo {volume} {78}},\ \bibinfo {pages}
  {195125} (\bibinfo {year} {2008})}\BibitemShut {NoStop}%
\bibitem [{\citenamefont {Kitaev}(2009)}]{azclass3}%
  \BibitemOpen
  \bibfield  {author} {\bibinfo {author} {\bibfnamefont {A.}~\bibnamefont
  {Kitaev}},\ }\bibfield  {title} {\bibinfo {title} {Periodic table for
  topological insulators and superconductors},\ }in\ \href@noop {} {\emph
  {\bibinfo {booktitle} {AIP conference proceedings}}},\ Vol.\ \bibinfo
  {volume} {1134}\ (\bibinfo {organization} {American Institute of Physics},\
  \bibinfo {year} {2009})\ pp.\ \bibinfo {pages} {22--30}\BibitemShut {NoStop}%
\bibitem [{\citenamefont {Hasan}\ and\ \citenamefont {Kane}(2010)}]{ti1}%
  \BibitemOpen
  \bibfield  {author} {\bibinfo {author} {\bibfnamefont {M.~Z.}\ \bibnamefont
  {Hasan}}\ and\ \bibinfo {author} {\bibfnamefont {C.~L.}\ \bibnamefont
  {Kane}},\ }\bibfield  {title} {\bibinfo {title} {Colloquium: topological
  insulators},\ }\href@noop {} {\bibfield  {journal} {\bibinfo  {journal}
  {Reviews of modern physics}\ }\textbf {\bibinfo {volume} {82}},\ \bibinfo
  {pages} {3045} (\bibinfo {year} {2010})}\BibitemShut {NoStop}%
\bibitem [{\citenamefont {Qi}\ and\ \citenamefont {Zhang}(2011)}]{ti2}%
  \BibitemOpen
  \bibfield  {author} {\bibinfo {author} {\bibfnamefont {X.-L.}\ \bibnamefont
  {Qi}}\ and\ \bibinfo {author} {\bibfnamefont {S.-C.}\ \bibnamefont {Zhang}},\
  }\bibfield  {title} {\bibinfo {title} {Topological insulators and
  superconductors},\ }\href@noop {} {\bibfield  {journal} {\bibinfo  {journal}
  {Reviews of modern physics}\ }\textbf {\bibinfo {volume} {83}},\ \bibinfo
  {pages} {1057} (\bibinfo {year} {2011})}\BibitemShut {NoStop}%
\bibitem [{\citenamefont {Chiu}\ \emph {et~al.}(2016)\citenamefont {Chiu},
  \citenamefont {Teo}, \citenamefont {Schnyder},\ and\ \citenamefont
  {Ryu}}]{ti3}%
  \BibitemOpen
  \bibfield  {author} {\bibinfo {author} {\bibfnamefont {C.-K.}\ \bibnamefont
  {Chiu}}, \bibinfo {author} {\bibfnamefont {J.~C.}\ \bibnamefont {Teo}},
  \bibinfo {author} {\bibfnamefont {A.~P.}\ \bibnamefont {Schnyder}},\ and\
  \bibinfo {author} {\bibfnamefont {S.}~\bibnamefont {Ryu}},\ }\bibfield
  {title} {\bibinfo {title} {Classification of topological quantum matter with
  symmetries},\ }\href@noop {} {\bibfield  {journal} {\bibinfo  {journal}
  {Reviews of Modern Physics}\ }\textbf {\bibinfo {volume} {88}},\ \bibinfo
  {pages} {035005} (\bibinfo {year} {2016})}\BibitemShut {NoStop}%
\bibitem [{\citenamefont {Dyson}(1962)}]{disorder1}%
  \BibitemOpen
  \bibfield  {author} {\bibinfo {author} {\bibfnamefont {F.~J.}\ \bibnamefont
  {Dyson}},\ }\bibfield  {title} {\bibinfo {title} {The threefold way.
  algebraic structure of symmetry groups and ensembles in quantum mechanics},\
  }\href@noop {} {\bibfield  {journal} {\bibinfo  {journal} {Journal of
  Mathematical Physics}\ }\textbf {\bibinfo {volume} {3}},\ \bibinfo {pages}
  {1199} (\bibinfo {year} {1962})}\BibitemShut {NoStop}%
\bibitem [{\citenamefont {Guhr}\ \emph {et~al.}(1998)\citenamefont {Guhr},
  \citenamefont {M{\"u}ller-Groeling},\ and\ \citenamefont
  {Weidenm{\"u}ller}}]{disorder2}%
  \BibitemOpen
  \bibfield  {author} {\bibinfo {author} {\bibfnamefont {T.}~\bibnamefont
  {Guhr}}, \bibinfo {author} {\bibfnamefont {A.}~\bibnamefont
  {M{\"u}ller-Groeling}},\ and\ \bibinfo {author} {\bibfnamefont {H.~A.}\
  \bibnamefont {Weidenm{\"u}ller}},\ }\bibfield  {title} {\bibinfo {title}
  {Random-matrix theories in quantum physics: common concepts},\ }\href@noop {}
  {\bibfield  {journal} {\bibinfo  {journal} {Physics Reports}\ }\textbf
  {\bibinfo {volume} {299}},\ \bibinfo {pages} {189} (\bibinfo {year}
  {1998})}\BibitemShut {NoStop}%
\bibitem [{\citenamefont {Ryu}\ \emph {et~al.}(2010)\citenamefont {Ryu},
  \citenamefont {Schnyder}, \citenamefont {Furusaki},\ and\ \citenamefont
  {Ludwig}}]{disorder3}%
  \BibitemOpen
  \bibfield  {author} {\bibinfo {author} {\bibfnamefont {S.}~\bibnamefont
  {Ryu}}, \bibinfo {author} {\bibfnamefont {A.~P.}\ \bibnamefont {Schnyder}},
  \bibinfo {author} {\bibfnamefont {A.}~\bibnamefont {Furusaki}},\ and\
  \bibinfo {author} {\bibfnamefont {A.~W.}\ \bibnamefont {Ludwig}},\ }\bibfield
   {title} {\bibinfo {title} {Topological insulators and superconductors:
  tenfold way and dimensional hierarchy},\ }\href@noop {} {\bibfield  {journal}
  {\bibinfo  {journal} {New Journal of Physics}\ }\textbf {\bibinfo {volume}
  {12}},\ \bibinfo {pages} {065010} (\bibinfo {year} {2010})}\BibitemShut
  {NoStop}%
\bibitem [{\citenamefont {Ryu}\ \emph {et~al.}(2012)\citenamefont {Ryu},
  \citenamefont {Moore},\ and\ \citenamefont {Ludwig}}]{disorder4}%
  \BibitemOpen
  \bibfield  {author} {\bibinfo {author} {\bibfnamefont {S.}~\bibnamefont
  {Ryu}}, \bibinfo {author} {\bibfnamefont {J.~E.}\ \bibnamefont {Moore}},\
  and\ \bibinfo {author} {\bibfnamefont {A.~W.}\ \bibnamefont {Ludwig}},\
  }\bibfield  {title} {\bibinfo {title} {Electromagnetic and gravitational
  responses and anomalies in topological insulators and superconductors},\
  }\href@noop {} {\bibfield  {journal} {\bibinfo  {journal} {Physical Review
  B—Condensed Matter and Materials Physics}\ }\textbf {\bibinfo {volume}
  {85}},\ \bibinfo {pages} {045104} (\bibinfo {year} {2012})}\BibitemShut
  {NoStop}%
\bibitem [{\citenamefont {Breuer}\ and\ \citenamefont
  {Petruccione}(2002)}]{open1}%
  \BibitemOpen
  \bibfield  {author} {\bibinfo {author} {\bibfnamefont {H.-P.}\ \bibnamefont
  {Breuer}}\ and\ \bibinfo {author} {\bibfnamefont {F.}~\bibnamefont
  {Petruccione}},\ }\href@noop {} {\emph {\bibinfo {title} {The theory of open
  quantum systems}}}\ (\bibinfo  {publisher} {OUP Oxford},\ \bibinfo {year}
  {2002})\BibitemShut {NoStop}%
\bibitem [{\citenamefont {Rivas}\ and\ \citenamefont {Huelga}(2012)}]{open2}%
  \BibitemOpen
  \bibfield  {author} {\bibinfo {author} {\bibfnamefont {A.}~\bibnamefont
  {Rivas}}\ and\ \bibinfo {author} {\bibfnamefont {S.~F.}\ \bibnamefont
  {Huelga}},\ }\href@noop {} {\emph {\bibinfo {title} {Open quantum
  systems}}},\ Vol.~\bibinfo {volume} {10}\ (\bibinfo  {publisher} {Springer},\
  \bibinfo {year} {2012})\BibitemShut {NoStop}%
\bibitem [{\citenamefont {Rivas}\ \emph {et~al.}(2013)\citenamefont {Rivas},
  \citenamefont {Viyuela},\ and\ \citenamefont {Martin-Delgado}}]{top_dens1}%
  \BibitemOpen
  \bibfield  {author} {\bibinfo {author} {\bibfnamefont {A.}~\bibnamefont
  {Rivas}}, \bibinfo {author} {\bibfnamefont {O.}~\bibnamefont {Viyuela}},\
  and\ \bibinfo {author} {\bibfnamefont {M.}~\bibnamefont {Martin-Delgado}},\
  }\bibfield  {title} {\bibinfo {title} {Density-matrix chern insulators:
  Finite-temperature generalization of topological insulators},\ }\href@noop {}
  {\bibfield  {journal} {\bibinfo  {journal} {Physical Review B—Condensed
  Matter and Materials Physics}\ }\textbf {\bibinfo {volume} {88}},\ \bibinfo
  {pages} {155141} (\bibinfo {year} {2013})}\BibitemShut {NoStop}%
\bibitem [{\citenamefont {Bardyn}\ \emph {et~al.}(2018)\citenamefont {Bardyn},
  \citenamefont {Wawer}, \citenamefont {Altland}, \citenamefont
  {Fleischhauer},\ and\ \citenamefont {Diehl}}]{top_dens2}%
  \BibitemOpen
  \bibfield  {author} {\bibinfo {author} {\bibfnamefont {C.-E.}\ \bibnamefont
  {Bardyn}}, \bibinfo {author} {\bibfnamefont {L.}~\bibnamefont {Wawer}},
  \bibinfo {author} {\bibfnamefont {A.}~\bibnamefont {Altland}}, \bibinfo
  {author} {\bibfnamefont {M.}~\bibnamefont {Fleischhauer}},\ and\ \bibinfo
  {author} {\bibfnamefont {S.}~\bibnamefont {Diehl}},\ }\bibfield  {title}
  {\bibinfo {title} {Probing the topology of density matrices},\ }\href@noop {}
  {\bibfield  {journal} {\bibinfo  {journal} {Physical Review X}\ }\textbf
  {\bibinfo {volume} {8}},\ \bibinfo {pages} {011035} (\bibinfo {year}
  {2018})}\BibitemShut {NoStop}%
\bibitem [{\citenamefont {Wawer}\ and\ \citenamefont
  {Fleischhauer}(2021{\natexlab{a}})}]{top_dens3}%
  \BibitemOpen
  \bibfield  {author} {\bibinfo {author} {\bibfnamefont {L.}~\bibnamefont
  {Wawer}}\ and\ \bibinfo {author} {\bibfnamefont {M.}~\bibnamefont
  {Fleischhauer}},\ }\bibfield  {title} {\bibinfo {title} {Chern number and
  berry curvature for gaussian mixed states of fermions},\ }\href@noop {}
  {\bibfield  {journal} {\bibinfo  {journal} {Physical Review B}\ }\textbf
  {\bibinfo {volume} {104}},\ \bibinfo {pages} {094104} (\bibinfo {year}
  {2021}{\natexlab{a}})}\BibitemShut {NoStop}%
\bibitem [{\citenamefont {Wawer}\ and\ \citenamefont
  {Fleischhauer}(2021{\natexlab{b}})}]{top_dens4}%
  \BibitemOpen
  \bibfield  {author} {\bibinfo {author} {\bibfnamefont {L.}~\bibnamefont
  {Wawer}}\ and\ \bibinfo {author} {\bibfnamefont {M.}~\bibnamefont
  {Fleischhauer}},\ }\bibfield  {title} {\bibinfo {title} {Z 2 topological
  invariants for mixed states of fermions in time-reversal invariant band
  structures},\ }\href@noop {} {\bibfield  {journal} {\bibinfo  {journal}
  {Physical Review B}\ }\textbf {\bibinfo {volume} {104}},\ \bibinfo {pages}
  {214107} (\bibinfo {year} {2021}{\natexlab{b}})}\BibitemShut {NoStop}%
\bibitem [{\citenamefont {Molignini}\ and\ \citenamefont
  {Cooper}(2023)}]{top_dens5}%
  \BibitemOpen
  \bibfield  {author} {\bibinfo {author} {\bibfnamefont {P.}~\bibnamefont
  {Molignini}}\ and\ \bibinfo {author} {\bibfnamefont {N.~R.}\ \bibnamefont
  {Cooper}},\ }\bibfield  {title} {\bibinfo {title} {Topological phase
  transitions at finite temperature},\ }\href@noop {} {\bibfield  {journal}
  {\bibinfo  {journal} {Physical Review Research}\ }\textbf {\bibinfo {volume}
  {5}},\ \bibinfo {pages} {023004} (\bibinfo {year} {2023})}\BibitemShut
  {NoStop}%
\bibitem [{\citenamefont {Huang}\ and\ \citenamefont
  {Diehl}(2024)}]{top_dens6}%
  \BibitemOpen
  \bibfield  {author} {\bibinfo {author} {\bibfnamefont {Z.-M.}\ \bibnamefont
  {Huang}}\ and\ \bibinfo {author} {\bibfnamefont {S.}~\bibnamefont {Diehl}},\
  }\bibfield  {title} {\bibinfo {title} {Mixed state topological order
  parameters for symmetry protected fermion matter},\ }\href@noop {} {\bibfield
   {journal} {\bibinfo  {journal} {arXiv preprint arXiv:2401.10993}\ }
  (\bibinfo {year} {2024})}\BibitemShut {NoStop}%
\bibitem [{\citenamefont {Mao}\ \emph {et~al.}(2024{\natexlab{a}})\citenamefont
  {Mao}, \citenamefont {Zhai},\ and\ \citenamefont {Yang}}]{top_dens7}%
  \BibitemOpen
  \bibfield  {author} {\bibinfo {author} {\bibfnamefont {L.}~\bibnamefont
  {Mao}}, \bibinfo {author} {\bibfnamefont {H.}~\bibnamefont {Zhai}},\ and\
  \bibinfo {author} {\bibfnamefont {F.}~\bibnamefont {Yang}},\ }\bibfield
  {title} {\bibinfo {title} {Probing topology of gaussian mixed states by the
  full counting statistics},\ }\href@noop {} {\bibfield  {journal} {\bibinfo
  {journal} {arXiv preprint arXiv:2402.15964}\ } (\bibinfo {year}
  {2024}{\natexlab{a}})}\BibitemShut {NoStop}%
\bibitem [{\citenamefont {Huang}\ and\ \citenamefont
  {Diehl}(2025)}]{top_dens8}%
  \BibitemOpen
  \bibfield  {author} {\bibinfo {author} {\bibfnamefont {Z.-M.}\ \bibnamefont
  {Huang}}\ and\ \bibinfo {author} {\bibfnamefont {S.}~\bibnamefont {Diehl}},\
  }\bibfield  {title} {\bibinfo {title} {Interaction-induced topological phase
  transition at finite temperature},\ }\href@noop {} {\bibfield  {journal}
  {\bibinfo  {journal} {Physical Review Letters}\ }\textbf {\bibinfo {volume}
  {134}},\ \bibinfo {pages} {053002} (\bibinfo {year} {2025})}\BibitemShut
  {NoStop}%
\bibitem [{\citenamefont {Bernard}\ and\ \citenamefont {LeClair}(2002)}]{bl1}%
  \BibitemOpen
  \bibfield  {author} {\bibinfo {author} {\bibfnamefont {D.}~\bibnamefont
  {Bernard}}\ and\ \bibinfo {author} {\bibfnamefont {A.}~\bibnamefont
  {LeClair}},\ }\bibfield  {title} {\bibinfo {title} {A classification of
  non-hermitian random matrices},\ }\href@noop {} {\bibfield  {journal}
  {\bibinfo  {journal} {Statistical Field Theories}\ ,\ \bibinfo {pages} {207}}
  (\bibinfo {year} {2002})}\BibitemShut {NoStop}%
\bibitem [{\citenamefont {Gong}\ \emph {et~al.}(2018)\citenamefont {Gong},
  \citenamefont {Ashida}, \citenamefont {Kawabata}, \citenamefont {Takasan},
  \citenamefont {Higashikawa},\ and\ \citenamefont {Ueda}}]{bl2}%
  \BibitemOpen
  \bibfield  {author} {\bibinfo {author} {\bibfnamefont {Z.}~\bibnamefont
  {Gong}}, \bibinfo {author} {\bibfnamefont {Y.}~\bibnamefont {Ashida}},
  \bibinfo {author} {\bibfnamefont {K.}~\bibnamefont {Kawabata}}, \bibinfo
  {author} {\bibfnamefont {K.}~\bibnamefont {Takasan}}, \bibinfo {author}
  {\bibfnamefont {S.}~\bibnamefont {Higashikawa}},\ and\ \bibinfo {author}
  {\bibfnamefont {M.}~\bibnamefont {Ueda}},\ }\bibfield  {title} {\bibinfo
  {title} {Topological phases of non-hermitian systems},\ }\href@noop {}
  {\bibfield  {journal} {\bibinfo  {journal} {Physical Review X}\ }\textbf
  {\bibinfo {volume} {8}},\ \bibinfo {pages} {031079} (\bibinfo {year}
  {2018})}\BibitemShut {NoStop}%
\bibitem [{\citenamefont {Kawabata}\ \emph {et~al.}(2019)\citenamefont
  {Kawabata}, \citenamefont {Shiozaki}, \citenamefont {Ueda},\ and\
  \citenamefont {Sato}}]{bl3}%
  \BibitemOpen
  \bibfield  {author} {\bibinfo {author} {\bibfnamefont {K.}~\bibnamefont
  {Kawabata}}, \bibinfo {author} {\bibfnamefont {K.}~\bibnamefont {Shiozaki}},
  \bibinfo {author} {\bibfnamefont {M.}~\bibnamefont {Ueda}},\ and\ \bibinfo
  {author} {\bibfnamefont {M.}~\bibnamefont {Sato}},\ }\bibfield  {title}
  {\bibinfo {title} {Symmetry and topology in non-hermitian physics},\
  }\href@noop {} {\bibfield  {journal} {\bibinfo  {journal} {Physical Review
  X}\ }\textbf {\bibinfo {volume} {9}},\ \bibinfo {pages} {041015} (\bibinfo
  {year} {2019})}\BibitemShut {NoStop}%
\bibitem [{\citenamefont {Zhou}\ and\ \citenamefont {Lee}(2019)}]{bl4}%
  \BibitemOpen
  \bibfield  {author} {\bibinfo {author} {\bibfnamefont {H.}~\bibnamefont
  {Zhou}}\ and\ \bibinfo {author} {\bibfnamefont {J.~Y.}\ \bibnamefont {Lee}},\
  }\bibfield  {title} {\bibinfo {title} {Periodic table for topological bands
  with non-hermitian symmetries},\ }\href@noop {} {\bibfield  {journal}
  {\bibinfo  {journal} {Physical Review B}\ }\textbf {\bibinfo {volume} {99}},\
  \bibinfo {pages} {235112} (\bibinfo {year} {2019})}\BibitemShut {NoStop}%
\bibitem [{\citenamefont {Liu}\ and\ \citenamefont {Chen}(2019)}]{bl5}%
  \BibitemOpen
  \bibfield  {author} {\bibinfo {author} {\bibfnamefont {C.-H.}\ \bibnamefont
  {Liu}}\ and\ \bibinfo {author} {\bibfnamefont {S.}~\bibnamefont {Chen}},\
  }\bibfield  {title} {\bibinfo {title} {Topological classification of defects
  in non-hermitian systems},\ }\href@noop {} {\bibfield  {journal} {\bibinfo
  {journal} {Physical Review B}\ }\textbf {\bibinfo {volume} {100}},\ \bibinfo
  {pages} {144106} (\bibinfo {year} {2019})}\BibitemShut {NoStop}%
\bibitem [{\citenamefont {Ashida}\ \emph {et~al.}(2020)\citenamefont {Ashida},
  \citenamefont {Gong},\ and\ \citenamefont {Ueda}}]{bl6}%
  \BibitemOpen
  \bibfield  {author} {\bibinfo {author} {\bibfnamefont {Y.}~\bibnamefont
  {Ashida}}, \bibinfo {author} {\bibfnamefont {Z.}~\bibnamefont {Gong}},\ and\
  \bibinfo {author} {\bibfnamefont {M.}~\bibnamefont {Ueda}},\ }\bibfield
  {title} {\bibinfo {title} {Non-hermitian physics},\ }\href@noop {} {\bibfield
   {journal} {\bibinfo  {journal} {Advances in Physics}\ }\textbf {\bibinfo
  {volume} {69}},\ \bibinfo {pages} {249} (\bibinfo {year} {2020})}\BibitemShut
  {NoStop}%
\bibitem [{\citenamefont {Diehl}\ \emph {et~al.}(2011)\citenamefont {Diehl},
  \citenamefont {Rico}, \citenamefont {Baranov},\ and\ \citenamefont
  {Zoller}}]{open_top4}%
  \BibitemOpen
  \bibfield  {author} {\bibinfo {author} {\bibfnamefont {S.}~\bibnamefont
  {Diehl}}, \bibinfo {author} {\bibfnamefont {E.}~\bibnamefont {Rico}},
  \bibinfo {author} {\bibfnamefont {M.~A.}\ \bibnamefont {Baranov}},\ and\
  \bibinfo {author} {\bibfnamefont {P.}~\bibnamefont {Zoller}},\ }\bibfield
  {title} {\bibinfo {title} {Topology by dissipation in atomic quantum wires},\
  }\href@noop {} {\bibfield  {journal} {\bibinfo  {journal} {Nature physics}\
  }\textbf {\bibinfo {volume} {7}},\ \bibinfo {pages} {971} (\bibinfo {year}
  {2011})}\BibitemShut {NoStop}%
\bibitem [{\citenamefont {Bardyn}\ \emph {et~al.}(2013)\citenamefont {Bardyn},
  \citenamefont {Baranov}, \citenamefont {Kraus}, \citenamefont {Rico},
  \citenamefont {{\.I}mamo{\u{g}}lu}, \citenamefont {Zoller},\ and\
  \citenamefont {Diehl}}]{open_top5}%
  \BibitemOpen
  \bibfield  {author} {\bibinfo {author} {\bibfnamefont {C.-E.}\ \bibnamefont
  {Bardyn}}, \bibinfo {author} {\bibfnamefont {M.~A.}\ \bibnamefont {Baranov}},
  \bibinfo {author} {\bibfnamefont {C.~V.}\ \bibnamefont {Kraus}}, \bibinfo
  {author} {\bibfnamefont {E.}~\bibnamefont {Rico}}, \bibinfo {author}
  {\bibfnamefont {A.}~\bibnamefont {{\.I}mamo{\u{g}}lu}}, \bibinfo {author}
  {\bibfnamefont {P.}~\bibnamefont {Zoller}},\ and\ \bibinfo {author}
  {\bibfnamefont {S.}~\bibnamefont {Diehl}},\ }\bibfield  {title} {\bibinfo
  {title} {Topology by dissipation},\ }\href@noop {} {\bibfield  {journal}
  {\bibinfo  {journal} {New Journal of Physics}\ }\textbf {\bibinfo {volume}
  {15}},\ \bibinfo {pages} {085001} (\bibinfo {year} {2013})}\BibitemShut
  {NoStop}%
\bibitem [{\citenamefont {Budich}\ \emph {et~al.}(2015)\citenamefont {Budich},
  \citenamefont {Zoller},\ and\ \citenamefont {Diehl}}]{open_top6}%
  \BibitemOpen
  \bibfield  {author} {\bibinfo {author} {\bibfnamefont {J.~C.}\ \bibnamefont
  {Budich}}, \bibinfo {author} {\bibfnamefont {P.}~\bibnamefont {Zoller}},\
  and\ \bibinfo {author} {\bibfnamefont {S.}~\bibnamefont {Diehl}},\ }\bibfield
   {title} {\bibinfo {title} {Dissipative preparation of chern insulators},\
  }\href@noop {} {\bibfield  {journal} {\bibinfo  {journal} {Physical Review
  A}\ }\textbf {\bibinfo {volume} {91}},\ \bibinfo {pages} {042117} (\bibinfo
  {year} {2015})}\BibitemShut {NoStop}%
\bibitem [{\citenamefont {Kawasaki}\ \emph {et~al.}(2022)\citenamefont
  {Kawasaki}, \citenamefont {Mochizuki},\ and\ \citenamefont
  {Obuse}}]{open_top1}%
  \BibitemOpen
  \bibfield  {author} {\bibinfo {author} {\bibfnamefont {M.}~\bibnamefont
  {Kawasaki}}, \bibinfo {author} {\bibfnamefont {K.}~\bibnamefont
  {Mochizuki}},\ and\ \bibinfo {author} {\bibfnamefont {H.}~\bibnamefont
  {Obuse}},\ }\bibfield  {title} {\bibinfo {title} {Topological phases
  protected by shifted sublattice symmetry in dissipative quantum systems},\
  }\href@noop {} {\bibfield  {journal} {\bibinfo  {journal} {Physical Review
  B}\ }\textbf {\bibinfo {volume} {106}},\ \bibinfo {pages} {035408} (\bibinfo
  {year} {2022})}\BibitemShut {NoStop}%
\bibitem [{\citenamefont {Altland}\ \emph {et~al.}(2021)\citenamefont
  {Altland}, \citenamefont {Fleischhauer},\ and\ \citenamefont
  {Diehl}}]{open_top2}%
  \BibitemOpen
  \bibfield  {author} {\bibinfo {author} {\bibfnamefont {A.}~\bibnamefont
  {Altland}}, \bibinfo {author} {\bibfnamefont {M.}~\bibnamefont
  {Fleischhauer}},\ and\ \bibinfo {author} {\bibfnamefont {S.}~\bibnamefont
  {Diehl}},\ }\bibfield  {title} {\bibinfo {title} {Symmetry classes of open
  fermionic quantum matter},\ }\href@noop {} {\bibfield  {journal} {\bibinfo
  {journal} {Physical Review X}\ }\textbf {\bibinfo {volume} {11}},\ \bibinfo
  {pages} {021037} (\bibinfo {year} {2021})}\BibitemShut {NoStop}%
\bibitem [{\citenamefont {Flynn}\ \emph {et~al.}(2021)\citenamefont {Flynn},
  \citenamefont {Cobanera},\ and\ \citenamefont {Viola}}]{open_top7}%
  \BibitemOpen
  \bibfield  {author} {\bibinfo {author} {\bibfnamefont {V.~P.}\ \bibnamefont
  {Flynn}}, \bibinfo {author} {\bibfnamefont {E.}~\bibnamefont {Cobanera}},\
  and\ \bibinfo {author} {\bibfnamefont {L.}~\bibnamefont {Viola}},\ }\bibfield
   {title} {\bibinfo {title} {Topology by dissipation: Majorana bosons in
  metastable quadratic markovian dynamics},\ }\href@noop {} {\bibfield
  {journal} {\bibinfo  {journal} {Physical Review Letters}\ }\textbf {\bibinfo
  {volume} {127}},\ \bibinfo {pages} {245701} (\bibinfo {year}
  {2021})}\BibitemShut {NoStop}%
\bibitem [{\citenamefont {Mao}\ \emph {et~al.}(2024{\natexlab{b}})\citenamefont
  {Mao}, \citenamefont {Yang},\ and\ \citenamefont {Zhai}}]{open_top3}%
  \BibitemOpen
  \bibfield  {author} {\bibinfo {author} {\bibfnamefont {L.}~\bibnamefont
  {Mao}}, \bibinfo {author} {\bibfnamefont {F.}~\bibnamefont {Yang}},\ and\
  \bibinfo {author} {\bibfnamefont {H.}~\bibnamefont {Zhai}},\ }\bibfield
  {title} {\bibinfo {title} {Symmetry-preserving quadratic lindbladian and
  dissipation driven topological transitions in gaussian states},\ }\href@noop
  {} {\bibfield  {journal} {\bibinfo  {journal} {Reports on Progress in
  Physics}\ }\textbf {\bibinfo {volume} {87}},\ \bibinfo {pages} {070501}
  (\bibinfo {year} {2024}{\natexlab{b}})}\BibitemShut {NoStop}%
\bibitem [{\citenamefont {S{\'a}}\ \emph
  {et~al.}(2020{\natexlab{a}})\citenamefont {S{\'a}}, \citenamefont {Ribeiro},\
  and\ \citenamefont {Prosen}}]{open_chaos1}%
  \BibitemOpen
  \bibfield  {author} {\bibinfo {author} {\bibfnamefont {L.}~\bibnamefont
  {S{\'a}}}, \bibinfo {author} {\bibfnamefont {P.}~\bibnamefont {Ribeiro}},\
  and\ \bibinfo {author} {\bibfnamefont {T.}~\bibnamefont {Prosen}},\
  }\bibfield  {title} {\bibinfo {title} {Complex spacing ratios: A signature of
  dissipative quantum chaos},\ }\href@noop {} {\bibfield  {journal} {\bibinfo
  {journal} {Physical Review X}\ }\textbf {\bibinfo {volume} {10}},\ \bibinfo
  {pages} {021019} (\bibinfo {year} {2020}{\natexlab{a}})}\BibitemShut
  {NoStop}%
\bibitem [{\citenamefont {Hamazaki}\ \emph {et~al.}(2020)\citenamefont
  {Hamazaki}, \citenamefont {Kawabata}, \citenamefont {Kura},\ and\
  \citenamefont {Ueda}}]{open_chaos2}%
  \BibitemOpen
  \bibfield  {author} {\bibinfo {author} {\bibfnamefont {R.}~\bibnamefont
  {Hamazaki}}, \bibinfo {author} {\bibfnamefont {K.}~\bibnamefont {Kawabata}},
  \bibinfo {author} {\bibfnamefont {N.}~\bibnamefont {Kura}},\ and\ \bibinfo
  {author} {\bibfnamefont {M.}~\bibnamefont {Ueda}},\ }\bibfield  {title}
  {\bibinfo {title} {Universality classes of non-hermitian random matrices},\
  }\href@noop {} {\bibfield  {journal} {\bibinfo  {journal} {Physical Review
  Research}\ }\textbf {\bibinfo {volume} {2}},\ \bibinfo {pages} {023286}
  (\bibinfo {year} {2020})}\BibitemShut {NoStop}%
\bibitem [{\citenamefont {Li}\ \emph {et~al.}(2021{\natexlab{a}})\citenamefont
  {Li}, \citenamefont {Prosen},\ and\ \citenamefont {Chan}}]{open_chaos3}%
  \BibitemOpen
  \bibfield  {author} {\bibinfo {author} {\bibfnamefont {J.}~\bibnamefont
  {Li}}, \bibinfo {author} {\bibfnamefont {T.}~\bibnamefont {Prosen}},\ and\
  \bibinfo {author} {\bibfnamefont {A.}~\bibnamefont {Chan}},\ }\bibfield
  {title} {\bibinfo {title} {Spectral statistics of non-hermitian matrices and
  dissipative quantum chaos},\ }\href@noop {} {\bibfield  {journal} {\bibinfo
  {journal} {Physical review letters}\ }\textbf {\bibinfo {volume} {127}},\
  \bibinfo {pages} {170602} (\bibinfo {year} {2021}{\natexlab{a}})}\BibitemShut
  {NoStop}%
\bibitem [{\citenamefont {Zanardi}\ and\ \citenamefont
  {Anand}(2021)}]{open_chaos4}%
  \BibitemOpen
  \bibfield  {author} {\bibinfo {author} {\bibfnamefont {P.}~\bibnamefont
  {Zanardi}}\ and\ \bibinfo {author} {\bibfnamefont {N.}~\bibnamefont
  {Anand}},\ }\bibfield  {title} {\bibinfo {title} {Information scrambling and
  chaos in open quantum systems},\ }\href@noop {} {\bibfield  {journal}
  {\bibinfo  {journal} {Physical Review A}\ }\textbf {\bibinfo {volume}
  {103}},\ \bibinfo {pages} {062214} (\bibinfo {year} {2021})}\BibitemShut
  {NoStop}%
\bibitem [{\citenamefont {Li}\ \emph {et~al.}(2021{\natexlab{b}})\citenamefont
  {Li}, \citenamefont {Prosen},\ and\ \citenamefont {Chan}}]{open_chaos5}%
  \BibitemOpen
  \bibfield  {author} {\bibinfo {author} {\bibfnamefont {J.}~\bibnamefont
  {Li}}, \bibinfo {author} {\bibfnamefont {T.}~\bibnamefont {Prosen}},\ and\
  \bibinfo {author} {\bibfnamefont {A.}~\bibnamefont {Chan}},\ }\bibfield
  {title} {\bibinfo {title} {Spectral statistics of non-hermitian matrices and
  dissipative quantum chaos},\ }\href@noop {} {\bibfield  {journal} {\bibinfo
  {journal} {Physical review letters}\ }\textbf {\bibinfo {volume} {127}},\
  \bibinfo {pages} {170602} (\bibinfo {year} {2021}{\natexlab{b}})}\BibitemShut
  {NoStop}%
\bibitem [{\citenamefont {Garc{\'\i}a-Garc{\'\i}a}\ \emph
  {et~al.}(2022)\citenamefont {Garc{\'\i}a-Garc{\'\i}a}, \citenamefont
  {S{\'a}},\ and\ \citenamefont {Verbaarschot}}]{open_chaos6}%
  \BibitemOpen
  \bibfield  {author} {\bibinfo {author} {\bibfnamefont {A.~M.}\ \bibnamefont
  {Garc{\'\i}a-Garc{\'\i}a}}, \bibinfo {author} {\bibfnamefont
  {L.}~\bibnamefont {S{\'a}}},\ and\ \bibinfo {author} {\bibfnamefont {J.~J.}\
  \bibnamefont {Verbaarschot}},\ }\bibfield  {title} {\bibinfo {title}
  {Symmetry classification and universality in non-hermitian many-body quantum
  chaos by the sachdev-ye-kitaev model},\ }\href@noop {} {\bibfield  {journal}
  {\bibinfo  {journal} {Physical Review X}\ }\textbf {\bibinfo {volume} {12}},\
  \bibinfo {pages} {021040} (\bibinfo {year} {2022})}\BibitemShut {NoStop}%
\bibitem [{\citenamefont {S{\'a}}\ \emph {et~al.}(2022)\citenamefont {S{\'a}},
  \citenamefont {Ribeiro},\ and\ \citenamefont {Prosen}}]{open_chaos7}%
  \BibitemOpen
  \bibfield  {author} {\bibinfo {author} {\bibfnamefont {L.}~\bibnamefont
  {S{\'a}}}, \bibinfo {author} {\bibfnamefont {P.}~\bibnamefont {Ribeiro}},\
  and\ \bibinfo {author} {\bibfnamefont {T.}~\bibnamefont {Prosen}},\
  }\bibfield  {title} {\bibinfo {title} {Lindbladian dissipation of
  strongly-correlated quantum matter},\ }\href@noop {} {\bibfield  {journal}
  {\bibinfo  {journal} {Physical Review Research}\ }\textbf {\bibinfo {volume}
  {4}},\ \bibinfo {pages} {L022068} (\bibinfo {year} {2022})}\BibitemShut
  {NoStop}%
\bibitem [{\citenamefont {Kulkarni}\ \emph {et~al.}(2022)\citenamefont
  {Kulkarni}, \citenamefont {Numasawa},\ and\ \citenamefont
  {Ryu}}]{open_chaos8}%
  \BibitemOpen
  \bibfield  {author} {\bibinfo {author} {\bibfnamefont {A.}~\bibnamefont
  {Kulkarni}}, \bibinfo {author} {\bibfnamefont {T.}~\bibnamefont {Numasawa}},\
  and\ \bibinfo {author} {\bibfnamefont {S.}~\bibnamefont {Ryu}},\ }\bibfield
  {title} {\bibinfo {title} {Lindbladian dynamics of the sachdev-ye-kitaev
  model},\ }\href@noop {} {\bibfield  {journal} {\bibinfo  {journal} {Physical
  Review B}\ }\textbf {\bibinfo {volume} {106}},\ \bibinfo {pages} {075138}
  (\bibinfo {year} {2022})}\BibitemShut {NoStop}%
\bibitem [{\citenamefont {Kawabata}\ \emph
  {et~al.}(2023{\natexlab{a}})\citenamefont {Kawabata}, \citenamefont
  {Kulkarni}, \citenamefont {Li}, \citenamefont {Numasawa},\ and\ \citenamefont
  {Ryu}}]{open_chaos9}%
  \BibitemOpen
  \bibfield  {author} {\bibinfo {author} {\bibfnamefont {K.}~\bibnamefont
  {Kawabata}}, \bibinfo {author} {\bibfnamefont {A.}~\bibnamefont {Kulkarni}},
  \bibinfo {author} {\bibfnamefont {J.}~\bibnamefont {Li}}, \bibinfo {author}
  {\bibfnamefont {T.}~\bibnamefont {Numasawa}},\ and\ \bibinfo {author}
  {\bibfnamefont {S.}~\bibnamefont {Ryu}},\ }\bibfield  {title} {\bibinfo
  {title} {Dynamical quantum phase transitions in sachdev-ye-kitaev
  lindbladians},\ }\href@noop {} {\bibfield  {journal} {\bibinfo  {journal}
  {Physical Review B}\ }\textbf {\bibinfo {volume} {108}},\ \bibinfo {pages}
  {075110} (\bibinfo {year} {2023}{\natexlab{a}})}\BibitemShut {NoStop}%
\bibitem [{\citenamefont {Kawabata}\ \emph
  {et~al.}(2023{\natexlab{b}})\citenamefont {Kawabata}, \citenamefont {Xiao},
  \citenamefont {Ohtsuki},\ and\ \citenamefont {Shindou}}]{open_chaos10}%
  \BibitemOpen
  \bibfield  {author} {\bibinfo {author} {\bibfnamefont {K.}~\bibnamefont
  {Kawabata}}, \bibinfo {author} {\bibfnamefont {Z.}~\bibnamefont {Xiao}},
  \bibinfo {author} {\bibfnamefont {T.}~\bibnamefont {Ohtsuki}},\ and\ \bibinfo
  {author} {\bibfnamefont {R.}~\bibnamefont {Shindou}},\ }\bibfield  {title}
  {\bibinfo {title} {Singular-value statistics of non-hermitian random matrices
  and open quantum systems},\ }\href@noop {} {\bibfield  {journal} {\bibinfo
  {journal} {PRX Quantum}\ }\textbf {\bibinfo {volume} {4}},\ \bibinfo {pages}
  {040312} (\bibinfo {year} {2023}{\natexlab{b}})}\BibitemShut {NoStop}%
\bibitem [{\citenamefont {Prosen}(2008)}]{3nd1}%
  \BibitemOpen
  \bibfield  {author} {\bibinfo {author} {\bibfnamefont {T.}~\bibnamefont
  {Prosen}},\ }\bibfield  {title} {\bibinfo {title} {Third quantization: a
  general method to solve master equations for quadratic open fermi systems},\
  }\href@noop {} {\bibfield  {journal} {\bibinfo  {journal} {New Journal of
  Physics}\ }\textbf {\bibinfo {volume} {10}},\ \bibinfo {pages} {043026}
  (\bibinfo {year} {2008})}\BibitemShut {NoStop}%
\bibitem [{\citenamefont {Prosen}(2010)}]{3nd2}%
  \BibitemOpen
  \bibfield  {author} {\bibinfo {author} {\bibfnamefont {T.}~\bibnamefont
  {Prosen}},\ }\bibfield  {title} {\bibinfo {title} {Spectral theorem for the
  lindblad equation for quadratic open fermionic systems},\ }\href@noop {}
  {\bibfield  {journal} {\bibinfo  {journal} {Journal of Statistical Mechanics:
  Theory and Experiment}\ }\textbf {\bibinfo {volume} {2010}},\ \bibinfo
  {pages} {P07020} (\bibinfo {year} {2010})}\BibitemShut {NoStop}%
\bibitem [{\citenamefont {Choi}(1975)}]{cj1}%
  \BibitemOpen
  \bibfield  {author} {\bibinfo {author} {\bibfnamefont {M.-D.}\ \bibnamefont
  {Choi}},\ }\bibfield  {title} {\bibinfo {title} {Completely positive linear
  maps on complex matrices},\ }\href@noop {} {\bibfield  {journal} {\bibinfo
  {journal} {Linear Alg. Appl.}\ }\textbf {\bibinfo {volume} {10}},\ \bibinfo
  {pages} {285} (\bibinfo {year} {1975})}\BibitemShut {NoStop}%
\bibitem [{\citenamefont {Jamio{\l}kowski}(1972)}]{cj2}%
  \BibitemOpen
  \bibfield  {author} {\bibinfo {author} {\bibfnamefont {A.}~\bibnamefont
  {Jamio{\l}kowski}},\ }\bibfield  {title} {\bibinfo {title} {Linear
  transformations which preserve trace and positive semidefiniteness of
  operators},\ }\href@noop {} {\bibfield  {journal} {\bibinfo  {journal} {Rep.
  Math. Phys.}\ }\textbf {\bibinfo {volume} {3}},\ \bibinfo {pages} {275}
  (\bibinfo {year} {1972})}\BibitemShut {NoStop}%
\bibitem [{\citenamefont {Barthel}\ and\ \citenamefont {Zhang}(2022)}]{eq1}%
  \BibitemOpen
  \bibfield  {author} {\bibinfo {author} {\bibfnamefont {T.}~\bibnamefont
  {Barthel}}\ and\ \bibinfo {author} {\bibfnamefont {Y.}~\bibnamefont
  {Zhang}},\ }\bibfield  {title} {\bibinfo {title} {Solving quasi-free and
  quadratic lindblad master equations for open fermionic and bosonic systems},\
  }\href@noop {} {\bibfield  {journal} {\bibinfo  {journal} {Journal of
  Statistical Mechanics: Theory and Experiment}\ }\textbf {\bibinfo {volume}
  {2022}},\ \bibinfo {pages} {113101} (\bibinfo {year} {2022})}\BibitemShut
  {NoStop}%
\bibitem [{\citenamefont {Song}\ \emph {et~al.}(2019)\citenamefont {Song},
  \citenamefont {Yao},\ and\ \citenamefont {Wang}}]{eq2}%
  \BibitemOpen
  \bibfield  {author} {\bibinfo {author} {\bibfnamefont {F.}~\bibnamefont
  {Song}}, \bibinfo {author} {\bibfnamefont {S.}~\bibnamefont {Yao}},\ and\
  \bibinfo {author} {\bibfnamefont {Z.}~\bibnamefont {Wang}},\ }\bibfield
  {title} {\bibinfo {title} {Non-hermitian skin effect and chiral damping in
  open quantum systems},\ }\href@noop {} {\bibfield  {journal} {\bibinfo
  {journal} {Physical review letters}\ }\textbf {\bibinfo {volume} {123}},\
  \bibinfo {pages} {170401} (\bibinfo {year} {2019})}\BibitemShut {NoStop}%
\bibitem [{Note1()}]{Note1}%
  \BibitemOpen
  \bibinfo {note} {We can always achieve this goal by properly add a phase
  factor $e^{i\theta }$ to $U$.}\BibitemShut {Stop}%
\bibitem [{Note2()}]{Note2}%
  \BibitemOpen
  \bibinfo {note} {The codes for numerical calculation are available at
  https://github.com/maol17/2503.10763}\BibitemShut {NoStop}%
\bibitem [{\citenamefont {S{\'a}}\ \emph
  {et~al.}(2020{\natexlab{b}})\citenamefont {S{\'a}}, \citenamefont {Ribeiro},\
  and\ \citenamefont {Prosen}}]{random3}%
  \BibitemOpen
  \bibfield  {author} {\bibinfo {author} {\bibfnamefont {L.}~\bibnamefont
  {S{\'a}}}, \bibinfo {author} {\bibfnamefont {P.}~\bibnamefont {Ribeiro}},\
  and\ \bibinfo {author} {\bibfnamefont {T.}~\bibnamefont {Prosen}},\
  }\bibfield  {title} {\bibinfo {title} {Spectral and steady-state properties
  of random liouvillians},\ }\href@noop {} {\bibfield  {journal} {\bibinfo
  {journal} {Journal of Physics A: Mathematical and Theoretical}\ }\textbf
  {\bibinfo {volume} {53}},\ \bibinfo {pages} {305303} (\bibinfo {year}
  {2020}{\natexlab{b}})}\BibitemShut {NoStop}%
\bibitem [{\citenamefont {Costa}\ \emph {et~al.}(2023)\citenamefont {Costa},
  \citenamefont {Ribeiro}, \citenamefont {De~Luca}, \citenamefont {Prosen},\
  and\ \citenamefont {S{\'a}}}]{random2}%
  \BibitemOpen
  \bibfield  {author} {\bibinfo {author} {\bibfnamefont {J.}~\bibnamefont
  {Costa}}, \bibinfo {author} {\bibfnamefont {P.}~\bibnamefont {Ribeiro}},
  \bibinfo {author} {\bibfnamefont {A.}~\bibnamefont {De~Luca}}, \bibinfo
  {author} {\bibfnamefont {T.}~\bibnamefont {Prosen}},\ and\ \bibinfo {author}
  {\bibfnamefont {L.}~\bibnamefont {S{\'a}}},\ }\bibfield  {title} {\bibinfo
  {title} {Spectral and steady-state properties of fermionic random quadratic
  liouvillians},\ }\href@noop {} {\bibfield  {journal} {\bibinfo  {journal}
  {SciPost Physics}\ }\textbf {\bibinfo {volume} {15}},\ \bibinfo {pages} {145}
  (\bibinfo {year} {2023})}\BibitemShut {NoStop}%
\bibitem [{\citenamefont {Kane}\ and\ \citenamefont {Mele}(2005)}]{kane2005z}%
  \BibitemOpen
  \bibfield  {author} {\bibinfo {author} {\bibfnamefont {C.~L.}\ \bibnamefont
  {Kane}}\ and\ \bibinfo {author} {\bibfnamefont {E.~J.}\ \bibnamefont
  {Mele}},\ }\bibfield  {title} {\bibinfo {title} {Z 2 topological order and
  the quantum spin hall effect},\ }\href@noop {} {\bibfield  {journal}
  {\bibinfo  {journal} {Phys. Rev. Lett.}\ }\textbf {\bibinfo {volume} {95}},\
  \bibinfo {pages} {146802} (\bibinfo {year} {2005})}\BibitemShut {NoStop}%
\bibitem [{\citenamefont {Wang}\ \emph {et~al.}()\citenamefont {Wang},
  \citenamefont {Wang},\ and\ \citenamefont {Wang}}]{Wang24}%
  \BibitemOpen
  \bibfield  {author} {\bibinfo {author} {\bibfnamefont {H.-R.}\ \bibnamefont
  {Wang}}, \bibinfo {author} {\bibfnamefont {Z.}~\bibnamefont {Wang}},\ and\
  \bibinfo {author} {\bibfnamefont {Z.}~\bibnamefont {Wang}},\ }\bibfield
  {title} {\bibinfo {title} {Non-bloch self-energy of dissipative interacting
  fermions},\ }\href@noop {} {\bibinfo  {journal}
  {https://arxiv.org/pdf/2411.13661}\ }\BibitemShut {NoStop}%
\end{thebibliography}%

\newpage

\end{document}